\documentclass{article}

\usepackage[english]{babel}
\usepackage[utf8x]{inputenc}
\usepackage[T1]{fontenc}
\usepackage{amsmath,amsxtra,amssymb,amsfonts,amsthm,amscd,array}
\usepackage{graphicx,latexsym,wrapfig}

\DeclareMathOperator*{\argmin}{arg\,min}
\usepackage{color}
\usepackage{mathtools}
\usepackage{float}
\usepackage{booktabs}
\usepackage{fancyhdr}
\usepackage{lscape}
\usepackage[colorinlistoftodos]{todonotes}
\usepackage[colorlinks=true, allcolors=blue]{hyperref}
\usepackage[authoryear,sort]{natbib}
\usepackage{appendix}
\usepackage{tikz}
\usetikzlibrary{matrix,chains,positioning,decorations.pathreplacing,arrows}

\newcommand{\crl}[1]{\ensuremath{ \left\{ #1 \right\} }}
\newcommand{\edg}[1]{\ensuremath{\! \left[ #1 \right] }}
\newcommand{\brak}[1]{\ensuremath{\left( #1 \right)}}

\newcommand{\be}{\begin{equation}}
\newcommand{\ee}{\end{equation}}
\newcommand{\bea}{\begin{eqnarray}}
\newcommand{\eea}{\end{eqnarray}}
\newcommand{\beas}{\begin{eqnarray*}}
\newcommand{\eeas}{\end{eqnarray*}}

\usepackage{xparse}
\usepackage{breqn}

\usepackage[top=2.8cm,bottom=2cm,left=3cm,right=3cm,marginparwidth=1.75cm]{geometry}

\theoremstyle{definition}

\newcommand{\R}{\mathbb{R}}

\newcommand{\p}{\mathbb{P}} 

\newcommand{\q}{\mathbb{Q}}
\newcommand{\E}{\mathbb{E}}

\title{Assessing asset-liability risk with neural networks}
\author{Patrick Cheridito, \quad John Ery\footnote{As SCOR Fellow, John Ery thanks SCOR for financial support.}, 
\quad Mario V. Wüthrich\\[2mm]
\hspace*{1mm} RiskLab, ETH Zurich}
\date{} 
\begin{document}
\maketitle

\begin{abstract}
We introduce a neural network approach for assessing the risk of a portfolio of 
assets and liabilities over a given time period. This requires a conditional valuation of 
the portfolio given the state of the world at a later time, a problem that is particularly challenging if 
the portfolio contains structured products or complex insurance contracts which do not admit closed form 
valuation formulas. We illustrate the method on different examples from banking and insurance. We 
focus on value-at-risk and expected shortfall, but the approach also works for other risk measures.\\[2mm]
\end{abstract}

\section{Introduction}
\label{sec:intro}

Different financial risk management problems require an assessment of the risk of a portfolio of 
assets and liabilities over a given time period. Banks, mutual funds and hedge funds usually calculate
value-at-risk numbers for different risks over a day, a week or longer time periods.
Under Solvency II, insurance companies have to calculate one-year value-at-risk, whereas 
the Swiss Solvency Test demands a computation of one-year expected shortfall. A determination 
of these risk figures requires a conditional valuation of the portfolio given the state of the world at 
a later time, called risk horizon. This is particularly challenging if the portfolio contains structured 
products or complicated insurance policies which do not admit closed form valuation formulas. 
In theory, the problem can be approached with nested simulation, which is a two-stage procedure.
In the outer stage, different scenarios are generated to model how the world could evolve until the risk horizon,
whereas in the inner stage, cash flows occurring after the risk horizon are simulated to estimate the value of the portfolio 
conditionally on each scenario; see, e.g., \cite{lee98}, \cite{lee03}, \cite{gordy08},
\cite{broadie} or \cite{brs2012}. While nested simulation can be shown to converge for 
increasing sample sizes, it is often too time-consuming to be useful in practical applications.
A more pragmatic alternative, usually used for short risk horizons, is the delta-gamma method, which approximates the 
portfolio loss with a second order Taylor polynomial; see, e.g., \cite{rouvinez}, \cite{britten} or \cite{duffiepan}. 
If first and second derivatives of the portfolio with respect to the underlying risk 
factors are accessible, the method is computationally efficient, but its accuracy depends on how well the 
second order Taylor polynomial approximates the true portfolio loss. Similarly, the replicating portfolio approach 
approximates future cashflows with a portfolio of liquid instruments that can be priced 
efficiently; see e.g., \cite{mario}, \cite{pelsser}, \cite{natolski} or \cite{filipovic}. 
Building on work on American option valuation \cite[see e.g.,][]{carriere, vanroy, longstaff}, 
\cite{broadie15} as well as \cite{bauer} have proposed
to regress future cash flows on finitely many basis functions depending on state variables known 
at the risk horizon. This gives good results in a number of applications. 
But typically, for it to work well, the basis functions have to be chosen well-adapted to the problem. 

In this paper we use a neural network approach to approximate the value of the portfolio at the risk horizon. 
Since our goal is to estimate tail risk measures such as value-at-risk and expected shortfall, we employ importance 
sampling when training the networks and estimating the risk figures. In addition, we try different 
regularization techniques and test the adequacy of the neural network approximations using the 
defining property of the conditional expectation. Neural networks have also been used 
for the calculation of solvency capital requirements by \cite{hejazi}, \cite{MAF} and \cite{scognamiglio}. 
But they all exclusively focus on value-at-risk, do not make use of importance sampling 
and apply neural networks in a slightly different way. \cite{hejazi} develop a neural network 
interpolation scheme within a nested simulation framework. \cite{MAF} and \cite{scognamiglio}
both use reduced-size nested Monte Carlo to generate training samples for the calibration 
of the neural network. Here, we directly regress future cash-flows on a neural network without
producing training samples. 

The remainder of the paper is organized as follows. In Section \ref{sec:alrisk}, we set up the 
underlying risk model, introduce the importance sampling distributions we use in the implementation 
of our method and recall how value-at-risk and expected shortfall can be estimated from a finite sample
of simulated losses. Section \ref{sec:nnapprox} discusses the training, validation and testing of our network 
approximation of the conditional valuation functional. In Section \ref{sec:ex} we illustrate our approach on 
three typical risk calculation problems: the calculation of risk capital for a single put option, 
a portfolio of different call and put options and a variable annuity contract with 
guaranteed minimum income benefit. Section \ref{sec:conclu} concludes.

\section{Asset-liability risk}
\label{sec:alrisk}

We denote the current time by $0$ and are interested in the value of a portfolio of assets and liabilities 
at a given risk horizon $\tau >0$.
Suppose all relevant events taking place until time $\tau$ are described by a $d$-dimensional random 
vector $X = (X_1, \dots, X_d)$ defined on a measurable space $(\Omega, {\cal F})$ that is equipped with two 
equivalent probability measures, $\p$ and $\q$. We think of $\p$ as the real-world probability measure and 
use it for risk measurement. $\q$ is a risk-neutral probability measure specifying the time-$\tau$ value of 
the portfolio through
\[
V = v(X) + \mathbb{E}^{\q} \edg{\sum_{i=1}^I \frac{N_{\tau}}{N_{t_i}} C_{t_i} \biggm| X},
\]
where $v$ is a measurable function from $\R^d$ to $\R$ describing the part of the portfolio 
whose value is directly given by the underlying risk factors $X_1, \dots, X_d$;
$C_{t_1}, \dots, C_{t_I}$ are random cash flows occurring at times $\tau < t_1 < \dots < t_I$;
and $N_{\tau}, N_{t_1}, \dots, N_{t_I}$ model the evolution of a numeraire process used for discounting.

Our goal is to compute $\rho(L)$ for a risk measure $\rho$ and the time $\tau$ net liability $L = -V$, 
which can be written as 
\[
L = \mathbb{E}^{\q} \edg{Y \mid X}
\]
for 
\[
Y = -v(X) - \sum_{i=1}^I \frac{N_{\tau}}{N_{t_i}} C_{t_i}.
\]
Our approach works for a wide class of risk measures $\rho$. But for the sake of concreteness, 
we concentrate on value-at-risk and expected shortfall. We follow the convention of \cite{emb} 
and define value-at-risk at a level $\alpha \in (0,1)$ such as $0.95$, $0.99$ or $0.995$, as the left $\alpha$-quantile
\be \label{defVaR}
{\rm VaR}_{\alpha}(L) := \min \crl{x \in \R : \p[L \le x] \ge \alpha} 
= \sup \crl{x \in \R : \p[L \ge x] > 1 - \alpha}
\ee
and expected shortfall as
\be \label{defES}
{\rm ES}_{\alpha}(L) := \frac{1}{1-\alpha} \int_{\alpha}^1 {\rm VaR}_u(L) du.
\ee
There exist different definitions of VaR and ES in the literature. But if the distribution function $F_L$ of $L$ 
is continuous and strictly increasing, they are all equivalent\footnote{More precisely, 
$\mbox{VaR}_{\alpha}$ is usually defined as an $\alpha$- or (1$-\alpha)$-quantile depending on 
whether it is applied to $L$ or $-L$. So up to the sign convention, all VaR definitions 
coincide if $F_L$ is strictly increasing. Similarly, different definitions of ES are 
equivalent if $F_L$ is continuous.}.

\subsection{Conditional expectations as minimizing functions}

Let $\p \otimes \q$ be the probability measure on ${\cal F}$ given by 
\[
\p \otimes \q[A] := \int_{\R^d} \q[A \mid X = x] \, \pi(dx),
\]
where $\pi$ is the distribution of $X$ under $\p$ and $\q[A \mid X=x]$ is a regular conditional 
version of $\q$ given $X$. We assume that $Y$ belongs to $L^2(\Omega, {\cal F}, \p \otimes \q)$. 
Then $L$ is of the form $L = l(X)$ for a measurable function $l \colon \R^d \to \R$ minimizing the 
mean squared distance
\be \label{msd}
\E^{\p \otimes \q} \edg{\brak{l(X)- Y}^2};
\ee
see, e.g., \cite{bru}. Note that $l$ minimizes \eqref{msd} if and only if 
\[
l(x) = \argmin_{u \in \R} \int_{\R} (u - y)^2 \q[Y \in dy \mid X = x] \quad \mbox{for $\pi$-almost all } x \in \R^d,
\]
where $\q[Y \in dy \mid X = x]$ is a regular conditional $\q$-distribution of $Y$ given $X$. 
This shows that $l$ is unique up to $\pi$-almost sure equality and can alternatively be characterized by 
\[
l(x) = \argmin_{u \in \R} \int_{\R} (u -y)^2 \q[Y \in dy \mid X = x] \quad \mbox{for $\nu$-almost all } x \in \R^d
\]
for any probability measure $\nu$ on $\R^d$ that is equivalent to $\pi$. In particular, if 
$\p^{\nu}$ is the probability measure on $\sigma(X)$ under which $X$ has distribution $\nu$,
and $Y$ is in $L^2(\Omega, {\cal F}, \p^{\nu} \otimes \q)$, then
$l$ can be determined by minimizing 
\be \label{msdnu}
\E^{\p^{\nu} \otimes \q} \edg{\brak{l(X) - Y}^2}
\ee 
instead of \eqref{msd}. Since we are going to approximate the expectation \eqref{msdnu} with averages of 
Monte Carlo samples, this will give us some flexibility in simulating $(X,Y)$. 

\subsection{Monte Carlo estimation of value-at-risk and expected shortfall}\label{sec:monte}

Let $X^1, \dots, X^n$ be independent $\p$-simulations of $X$.
By $X^{(1)}, \dots, X^{(n)}$ we denote the same sample reordered so that 
\[
L^{(1)} = l(X^{(1)}) \ge \dots \ge L^{(n)} = l(X^{(n)}).
\]
To obtain $\mathbb{P}$-simulation estimates of $\mbox{VaR}_{\alpha}(L)$ and $\mbox{ES}_{\alpha}(L)$, 
we apply the VaR and ES definitions \eqref{defVaR}--\eqref{defES} to the empirical measure 
$\frac{1}{n} \sum_{i=1}^n \delta_{L^{(i)}}$. This yields
\be \label{PEst}
\widehat{\mbox{VaR}}_{\alpha}(n) := L^{(j)} \quad \mbox{and} \quad
\widehat{\mbox{ES}}_{\alpha}(n) := \frac{1}{1-\alpha} \sum_{i = 1}^{j-1} \frac{L^{(i)}}{n}
+ \brak{1 - \frac{j-1}{(1-\alpha)n}} L^{(j)},
\ee
where 
\[
j = \min \crl{i \in \crl{1, \dots, n} : i/n > 1-\alpha};
\]
see also \cite{emb}. It is well known that if $F_L$ is differentiable at $\mbox{VaR}_{\alpha}(L)$ with 
$F'_L(\mbox{VaR}_{\alpha}(L)) > 0$, then $\widehat{\mbox{VaR}}_{\alpha}(n)$ is an asymptotically 
normal consistent estimator of order $n^{-1/2}$; see e.g., 
\cite{david}. If in addition, $L$ is square-integrable, the same is true for $\widehat{\mbox{ES}}_{\alpha}(n)$; see
\cite{Zwingmann}.

\subsection{Importance sampling}\label{sec:tilt}

Assume now that $X$ is of the form $X = h(Z)$ for a transformation $h \colon \R^k \to \R^d$
and a $k$-dimensional random vector $Z$ with density $f \colon \R^k \to \R_+$. To give more weight 
to important outcomes, we introduce a second density $g$ on $\R^k$ satisfying $g(z) > 0$ for all 
$z \in \mathbb{R}^k$, where $f(z) > 0$. Let $Z^g$ be a $k$-dimensional random vector with density $g$ and 
$Z^{g,1}, \dots, Z^{g,n}$ independent simulations of $Z^g$. We reorder the simulations such that
\[
L^{g,(1)} = l\circ h(Z^{g,(1)}) \ge \dots \ge L^{g,(n)} = l \circ h(Z^{g,(n)})
\]
and consider the random weights 
\[
w^{(i)} = \frac{f(Z^{g,(i)})}{n g(Z^{g,(i)})}.
\]
Since $f(Z^g)/g(Z^g)$ is integrable, one obtains from the law of large numbers that 
for every threshold $x \in \mathbb{R}$, 
\be \label{gest}
\sum_{i=1}^n 1_{\crl{l \circ h(Z^{g,(i)}) \ge x}} w^{(i)}
\ee
is an unbiased consistent estimator of the exceedance probability
\[
\p [L \ge x] = \p \edg{l \circ h(Z) \ge x} = \E \edg{1_{\crl{l \circ h(Z^g) \ge x}} \frac{f(Z^g)}{g(Z^g)}}.
\]
If in addition, $g$ can be chosen so that $f(Z^g)/g(Z^g)$ is square-integrable, it follows from the 
central limit theorem that \eqref{gest} is asymptotically normal with a standard deviation 
of order $n^{-1/2}$. In any case, $\sum_{i=1}^n w^{(i)}$ converges to $1$ and the random measure 
$\sum_{i=1}^n w^{(i)} \delta_{l \circ h(Z^{g,(i)})}$ approximates the distribution of $L$.
Accordingly, we adapt the $\p$-simulation estimators \eqref{PEst} by replacing $i/n$ with 
$\sum_{m=1}^i w^{(m)}$. This yields the $g$-simulation estimates 
\[
\widehat{\mbox{VaR}}^g_{\alpha}(n) := L^{g,(j)} \quad \mbox{and} \quad
\widehat{\mbox{ES}}^g_{\alpha}(n) := \frac{1}{1-\alpha} \sum_{i = 1}^{j-1} w^{(i)} L^{g,(i)}
+ \brak{1 - \frac{1}{1- \alpha} \sum_{i=1}^{j-1} w^{(i)}} L^{g,(j)},
\]
where
\[
j = \min \crl{i \in \crl{1, \dots, n} : \sum_{m=1}^i w^{(m)} > 1-\alpha}.
\]
If the $\alpha$-quantile $l_{\alpha}$ of $L = l \circ h(Z)$ were known, 
the exceedance probability $\p \edg{L \ge l_{\alpha}}$ could be estimated 
by means of \eqref{gest} with $x = l_{\alpha}$. To make the procedure efficient, one would try to find a 
density $g$ on $\mathbb{R}^k$ from which it is easy to sample and such that the variance
\[
\mbox{Var} \brak{1_{\crl{l \circ h(Z^g) \ge l_{\alpha}}} \frac{f(Z^g)}{g(Z^g)}}
= \mathbb{E} \edg{1_{\crl{l \circ h(Z^g) \ge l_{\alpha}}} \frac{f(Z^g)^2}{g(Z^g)^2}}
- \mathbb{E} \edg{1_{\crl{l \circ h(Z^g) \ge l_{\alpha}}} \frac{f(Z^g)}{g(Z^g)}}^2,
\]
becomes as small as possible. Since 
\[
\mathbb{E} \edg{1_{\crl{l \circ h(Z^g) \ge l_{\alpha}}} \frac{f(Z^g)}{g(Z^g)}}
= \mathbb{E} \edg{1_{\crl{l \circ h(Z) \ge l_{\alpha}}}} = \mathbb{P} [L \ge l_{\alpha}]
\]
does not depend on $g$, it can be seen that $g$ is a good importance sampling (IS) density if
\begin{equation} \label{ming}
\mathbb{E} \edg{1_{\crl{l \circ h (Z^g) \ge l_{\alpha}}} \frac{f(Z^g)^2}{g(Z^g)^2}}
= \mathbb{E} \edg{1_{\crl{l \circ h (Z) \ge l_{\alpha}}} \frac{f(Z)}{g(Z)}}
\end{equation}
is small. We use the same criterion as a basis to find a good IS density for estimating
$\mbox{VaR}_{\alpha}$ and $\mbox{ES}_{\alpha}$; see e.g., \cite{glasserman} 
for more background on importance sampling.

\subsection{The case of an underlying multivariate normal distribution}\label{sec:impsamp}
\label{Subsec:nis}

In many applications $X$ can be modeled as a deterministic transformation of a random vector 
with a multivariate normal distribution; see e.g., the examples in Section \ref{sec:ex} below. 
In this case, it can be written as
\be \label{Xform}
X = u(AZ)
\ee for a function $u \colon \mathbb{R}^p \to \mathbb{R}^d$, a $p \times k$-matrix $A$ and 
a $k$-dimensional random vector $Z$ with a standard normal 
density $f$. To keep the problem tractable, we look for a suitable IS density $g$ in the class of
$N_k(m, I_k)$-densities $f_m$ with different mean vectors $m \in \R^k$ and covariance matrix 
equal to the $k$-dimensional identity matrix $I_k$. To determine a 
good choice of $m$, let $v \in \mathbb{R}^p$ be the vector with components 
\[
v_i = \begin{cases} 
1 & \mbox{ if $l \circ u$ is increasing in } y_i\\
-1 & \mbox{ if $l \circ u$ is decreasing in } y_i\\
0 & \mbox{ else}.
\end{cases}
\]
If $A^T v=0$, we choose $m = 0$. Otherwise, $v^T AZ/\|A^T v\|_2$ is standard normal. 
Denote its $\alpha$-quantile by $z_{\alpha}$. Then $Z$ falls into the region
\[
D = \crl{z \in \mathbb{R}^k : v^T A z/\|A^T v\|_2 \ge z_{\alpha}}
\]
with probability $1 - \alpha$, and $l \circ u(Az)$ tends to be large for $z \in D$. We choose $m$ as the maximizer of
$f(z)$ subject to $z \in D$, which leads to
\[
m = \frac{A^T v}{\|A^T v\|_2} z_{\alpha}.
\]
It is easy to see that this yields 
\[
f(z) < f_m(z) \quad \mbox{for all } z \in D.
\]
So, if $D$ is sufficiently similar to the region $\crl{z \in \mathbb{R}^k : l \circ u(Az) \ge l_{\alpha}}$, 
it can be seen from \eqref{ming} that this choice of IS density will yield a reduction in variance. 

\section{Neural network approximation}\label{sec:nnapprox}

Usually, the distribution of the risk factor vector $X$ is assumed to be known. For instance, in all 
our examples in Section \ref{sec:ex}, they are of the form \eqref{Xform}. On the other hand, in 
many real-world applications, there is no 
closed form expression for the loss function $l \colon \R^d \to \R$ mapping $X$ to $L$. Therefore, we
approximate $L = l(X)$ with $l^{\theta}(X)$ for a neural 
network $l^{\theta} \colon \R^d \to \R$; see e.g., \cite{goodfellow}. We concentrate on feedforward 
neural networks of the form 
\[
l^{\theta}  = \psi \circ a^{\theta}_J \circ \varphi \circ a^{\theta}_J \circ \dots \circ \varphi \circ a^{\theta}_1,
\]
where 
\begin{itemize}
\item 
$q_0 = d$, $q_{J}= 1$, and $q_1, \dots, q_{J-1}$ are the numbers of nodes in the hidden layers $1, \dots, J-1;$
\item 
$a^{\theta}_j$ are affine functions of the form $a^{\theta}_j(x) = A_j x + b_j$ for matrices 
$A_j \in \R^{q_j \times q_{j-1}}$ and vectors $b_j \in \R^{q_j},$ for $j=1,\ldots, J;$
\item
$\varphi$ is a non-linear activation function used in the hidden layers and applied component-wise. In the examples 
in Section \ref{sec:ex} we choose $\varphi = \tanh;$
\item
$\psi$ is the final activation function. For a portfolio of assets and liabilities a natural choice is 
$\psi = \mbox{id}$. To keep the presentation simple, we will consider pure liability portfolios with
loss $L > 0$ in all our examples below. Accordingly, we choose $\psi = \exp$.
\end{itemize}
The parameter vector $\theta$ consists of the components of the matrices $A_j$ and vectors $b_j$, 
$j = 1, \dots, J$. So it lives in $\mathbb{R}^q$ for $q = \sum_{j=1}^J q_j(q_{j-1} + 1)$.
It remains to determine the architecture of the network (that is, $J$ and $q_1, \dots, q_{J-1}$)
and to calibrate $\theta \in \mathbb{R}^q$. Then VaR and ES figures can be 
estimated as described in Section \ref{sec:alrisk} by simulating $l^{\theta}(X)$.

\subsection{Training and validation} 
\label{sec:trainval}

In a first step we take the network architecture ($J$ and $q_1, \dots, q_{J-1})$ as given and try to 
find a minimizer of $\theta \mapsto \E \edg{(l^{\theta}(X) - Y)^2}$, where the expectation is 
either with respect to $\p \otimes \q$ or $\p^{\nu} \otimes \q$ for an IS distribution $\nu$
on $\R^d$. To do that we simulate realizations $(X^m, Y^m)$, $m = 1, \dots, M_1 + M_2$, of $(X,Y)$
under the corresponding distribution. The first $M_1$ simulations are used for training and the other $M_2$ 
for validation. More precisely, we employ a stochastic gradient descent method to minimize the Monte 
Carlo approximation based on the training samples
\be \label{trainingerror}
\frac{1}{M_1} \sum_{m =1}^{M_1} \brak{l^{\theta}(X^m) - Y^m}^2
\ee
of $\mathbb{E} \edg{(l^{\theta}(X) - Y)}$. At the same time we use the validation samples to check whether 
\be \label{valerror}
\frac{1}{M_2} \sum_{m = M_1 + 1}^{M_1 + M_2} \brak{l^{\theta}(X^m) - Y^m}^2
\ee
is decreasing as we are updating $\theta$. 

\subsubsection{Regularization through tree structures}
\label{subsec:tree}

If the number $q$ of parameters is large, one needs to be careful not to overfit the 
neural network. For instance, in the extreme case, the 
network could be so flexible that it can bring \eqref{trainingerror} down to zero even in cases where 
the true conditional expectation $l(X) = \E^{\q}[Y \mid X]$ is not equal to $Y$. To prevent this, one can 
generate the training samples by first simulating $N_1$ realizations $X^i$ of $X$ and then for 
every $X^i$, drawing $N_2$ simulations $Y^{i,j}$ from the conditional distribution of $Y$ given $X^i$.
In the simple example of Section \ref{Subsec:opt}, we chose $N_2 = 1$. In Sections 
\ref{Subsec:20opt} and \ref{Subsec:va} we used $N_2 = 5$.

\subsubsection{Stochastic gradient descent} \label{sec:opti}

In principle, one can use any numerical method to minimize \eqref{trainingerror}. But 
stochastic gradient descent methods have proven to work well for neural networks. We refer to
\cite{ruder} for an overview of different (stochastic) gradient descent algorithms. Here, we 
randomly\footnote{If the training data is generated according to a tree 
structure as in Section \ref{subsec:tree}, one can either group the simulations 
$(X^i, Y^{i,j})$, $i = 1, \dots, N_1$, $j = 1, \dots, N_2$, so that pairs with the same $X^i$-component 
stay together or not. In our implementations, both methods gave similar results.} split the $M_1$ training 
samples into $b$ mini-batches of size $B$. Then we update $\theta$ based on the $\theta$-gradients of 
\[
\frac{1}{B} \sum_{m = (i-1)B + 1}^{iB} \brak{l^{\theta}(X^m) - Y^m}^2, \quad i = 1, \dots, b.
\]
We use batch normalization and Adam updating with the default values from \texttt{TensorFlow}.

After $b$ gradient steps, all of the training data have been used once and the first epoch is complete. 
For further epochs, we reshuffle the training data, form new mini-batches and perform $b$ more 
gradient steps. The procedure is repeated until the training error \eqref{trainingerror} stops to decrease 
or the validation error \eqref{valerror} starts to increase.

\subsubsection{Initialization}
\label{subsec:ini}

We follow standard practice and initialize the parameters of the network randomly. 
The final operation of the network is 
\[
x \mapsto \psi(A_J x + b_J).
\]
Since the network tries to approximate $Y$, and in all our examples below we use $\psi = \exp$, 
we initialize the last bias as
\[
b^0_J = \log\left(\frac{1}{M_1} \sum_{m=1}^{M_1} Y^m\right).
\]
For the other parameters we use Xavier initialization; see \cite{bengio}.

\subsection{Backtesting the network approximation} 
\label{subsec:backtest}

After having determined an approximate minimizer $\theta \in \R^q$ for a given network architecture, 
one can test the adequacy of the approximation $l^{\theta}(X)$ 
of the true conditional expectation $l(X) = \mathbb{E}^{\q}[Y \mid X]$. 
The quality of the approximation depends on different aspects:

\begin{itemize}
\item[(i)] {\sl Generalization error:}\\
The true conditional expectation $\E^{\q} \edg{Y \mid X}$ is of the form $l(X)$ for the 
unique\footnote{More precisely, uniqueness holds if functions are identified that
agree $\pi$-almost surely.} measurable function $l \colon \R^d \to \R$ minimizing the 
mean squared distance $\E \edg{\brak{l(X) - Y}^2}$.
To approximate $l$ we choose a network architecture and try to find a 
$\theta \in \mathbb{R}^q$ that minimizes the empirical squared distance
\be \label{discproblem}
\frac{1}{M_1} \sum_{m = 1}^{M_1} \brak{l^{\theta}(X^m) - Y^m}^2.
\ee
But if the samples $(X^m, Y^m),$ $m = 1, \dots, M_1$, do not represent the distribution of $(X,Y)$ well, 
\eqref{discproblem} might not be a good approximation of the true 
expectation $\E \edg{\brak{l^{\theta}(X) - Y}^2}$.

\item[(ii)]
{\sl Numerical minimization method:}\\
The minimization of \eqref{discproblem} usually is a complex problem, and one has to
employ a numerical method to find an approximate solution $\theta$. The 
quality of $l^{\theta}(X)$ will depend on the performance of the numerical method being used.

\item[(iii)]
{\sl Network architecture:}\\ It is well known that feedforward neural networks with one hidden layer have 
the universal approximation property; see e.g. \cite{cybenko}, \cite{hornik} or \cite{leshno}. That is, 
they can approximate any continuous function uniformly on compacts to any degree of 
accuracy if the activation function is of a suitable form and 
the hidden layers contain sufficiently many nodes. As a consequence, 
$\mathbb{E} \edg{(l^{\theta}(X) - l(X))^2}$ can be made arbitrarily small if 
the hidden layer is large enough and $\theta$ is chosen appropriately. 
However, we do not know in advance how many nodes we need. And moreover, 
feedforward neural networks with two or more hidden layers have shown to yield better
results in different applications.
\end{itemize}

Since we simulate from an underlying model, we are able to choose the size $M_1$ of the training 
sample large and train extensively. In addition, for any given 
network architecture, we also evaluate the empirical squared distance \eqref{discproblem} 
on the validation set $(X^m, Y^m)$, $m = M_1 +1, \dots, M_2$. So we suppose the
generalization error is small and our numerical method finds a good 
approximate minimizer $\theta$ of \eqref{discproblem}. But since we do not know whether a given network 
architecture is flexible enough to provide a good approximation to the true loss function $l$, we 
test for each trained network whether it satisfies the defining properties of a conditional expectation.

The loss function $l \colon \R^d \to \R$ is characterized by 
\begin{equation} \label{B}
\mathbb{E}[l(X) \xi(X)] = \mathbb{E}[Y \xi(X)] 
\end{equation}
for all measurable functions $\xi \colon \R^d \to \R$ such that $Y \xi(X)$ is integrable.
Ideally, we would like $l^{\theta}$ to satisfy the same condition. But there will be an approximation 
error, and \eqref{B} cannot be checked for all measurable functions $\xi \colon \R^d \to \R$
satisfying the integrability condition. Therefore, we select finitely many measurable subsets 
$B_i \subseteq \R^d$, $i = 1, \dots, I$. Then we generate $M_3$ more samples $(X^m, Y^m)$ 
of $(X,Y)$ and test whether the differences 

\begin{itemize}
\item[(a)]
$\sum_{m = M_2 +1}^{M_1 + M_2 + M_3} \brak{ l^{\theta}(X^m) - Y^m}/M_3$
\item[(b)]  
$\sum_{m = M_2 +1}^{M_1 + M_2 + M_3} \brak{l^{\theta}(X^m) -Y^m}  l^{\theta}(X^m)/M_3 $
\item[(c)] $\sum_{m = M_2 + 1}^{M_1 + M_2 + M_3} \brak{l^{\theta}(X^m) - Y^m}1_{B_i}(X^m)/M_3$ 
\end{itemize}
are sufficiently close to zero. If this is not the case, we change the network architecture and train again.

\section{Examples} \label{sec:ex}

As examples, we study three different risk assessment problems from banking and insurance. For comparison 
we generated realizations $(X^m, Y^m)$ of $(X,Y)$ under $\p \otimes \q$ as well as 
$\p^{\nu} \otimes \q$, where $\nu$ is the IS distribution on $\R^d$ obtained by 
changing the distribution of $X$. In all our examples, $X$ has a transformed
normal distribution as in Section \ref{Subsec:nis}. In each case we
used 1.5 million simulations with mini-batches of size 10,000 for training, 500,000 simulations for validation 
and 500,000 for backtesting. After training and testing the network, we simulated another 500,000 
realizations of $X$, once under $\p$ and then under $\p^{\nu}$,
to estimate $\mbox{VaR}_{99.5\%}(L)$ and $\mbox{ES}_{99\%}(L)$.

We implemented the algorithms in Python. To train the networks we used the \texttt{TensorFlow} package.

\subsection{Single put option} 
\label{Subsec:opt}

As a first example, we consider a liability consisting of a single put option with strike price $K = 100$ 
and maturity $T = 1/3$ on an underlying asset starting from $s_0 = 100$ and evolving according to 
\[
dS_t = \mu S_t dt + \sigma S_t dW^{\p}_t = r S_t dt + \sigma S_t dW^{\q}_t
\]
for an interest rate $r = 1\%$, a drift $\mu = 5\%$, a volatility $\sigma = 20\%$, a $\p$-Brownian motion 
$W^{\p}$ and a $\q$-Brownian motion $W^{\q}$. As risk horizon we choose $\tau = 1/52$.
The time $\tau$-value of this liability is 
\[
L =e^{- r(T-\tau)} \E^{\q} \edg{(K- S_T)^+ \mid S_{\tau}}.
\]
Using It\^{o}'s formula, one can write
\[
S_{\tau} = s_0 \exp \brak{(\mu - \sigma^2/2) \tau + \sigma \sqrt{\tau} Z} \quad \mbox{and} \quad 
S_T = S_{\tau} \exp \brak{(r - \sigma^2/2)(T-\tau) + \sigma \sqrt{T-\tau} \, V}, 
\]
where $Z$ and $V$ are two independent standard normal random variables under $\p \otimes \q$.
It is well-known that $L$ is of the form $P(S_{\tau}, r, \sigma, T- \tau)$, where $P$ is the Black--Scholes 
formula for a put option. This allows us to calculate reference 
values for $\mbox{VaR}_{\alpha}(L)$ and $\mbox{ES}_{\alpha}(L)$.

To train the neural network approximation of $L$, we simulated realizations of the pair $(X,Y)$ for 
$X=S_\tau$ and $Y = e^{-r(T-\tau)}(K-S_T)^+$. For comparison, we first 
simulated according to $\p \otimes \q$ and then according to $\p^{\nu} \otimes \q$
for an IS distribution $\p^{\nu}$. Clearly, $L$ is decreasing in $Z$. Therefore, we chose $\p^{\nu}$ 
so as to make $Z$ normally distributed with mean equal to the $1-\alpha$-quantile of a standard normal
and variance 1. Since $x \mapsto P(x, r, \sigma , T- \tau)$ is a simple one-dimensional function, we selected
a network with a single hidden layer containing 5 nodes. This is sufficient to approximate $P$ and
faster to train than a network with more hidden layers. We did not use the tree structure 
of Section \ref{subsec:tree} for training (that is, $N_2 = 1$) and trained the network over 40 epochs.

\begin{figure}[ht!] 
\centering
\begin{minipage}{.5\textwidth}
  \centering
  \includegraphics[width=.85\textwidth]{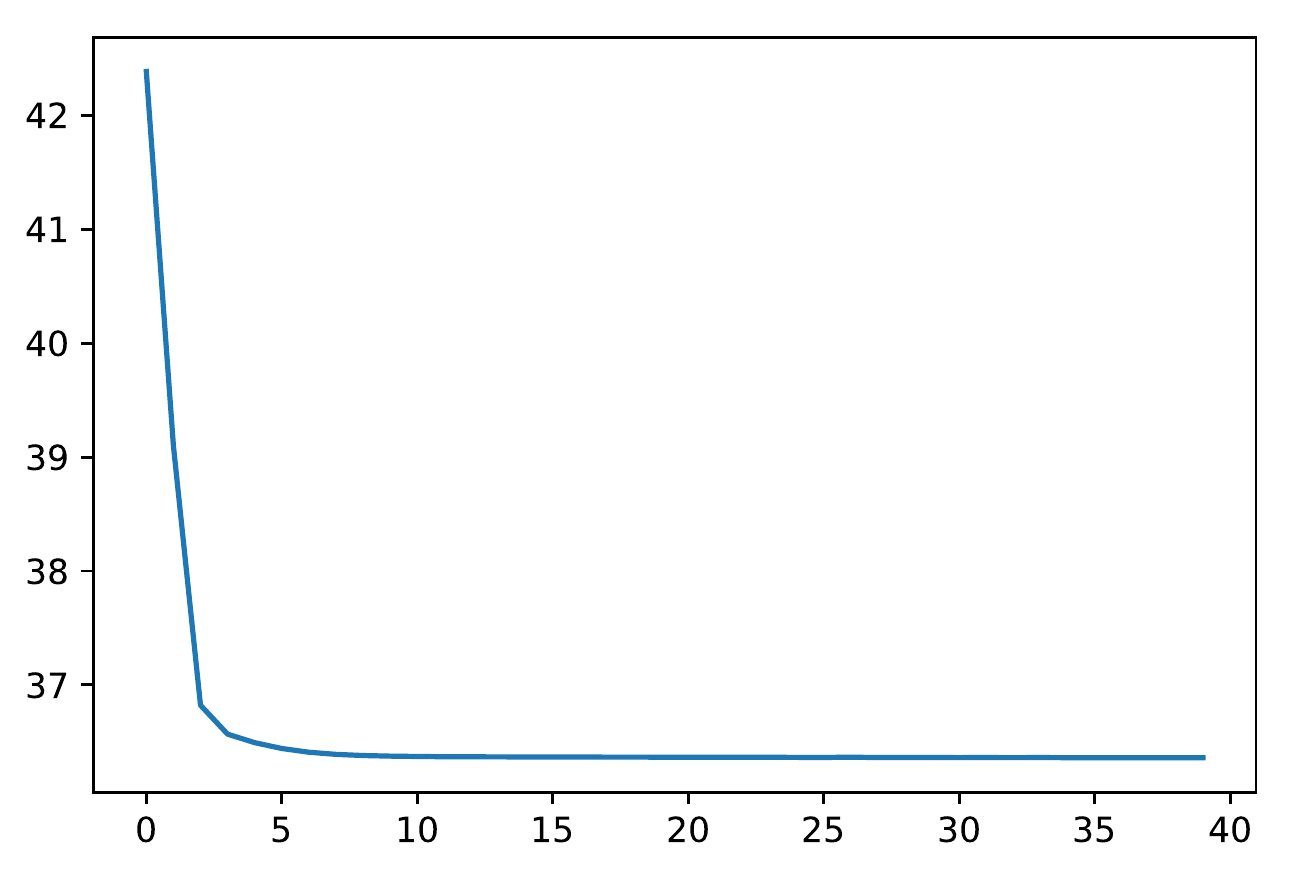}
\end{minipage}%
\begin{minipage}{.5\textwidth}
  \centering
  \includegraphics[width=.85\textwidth]{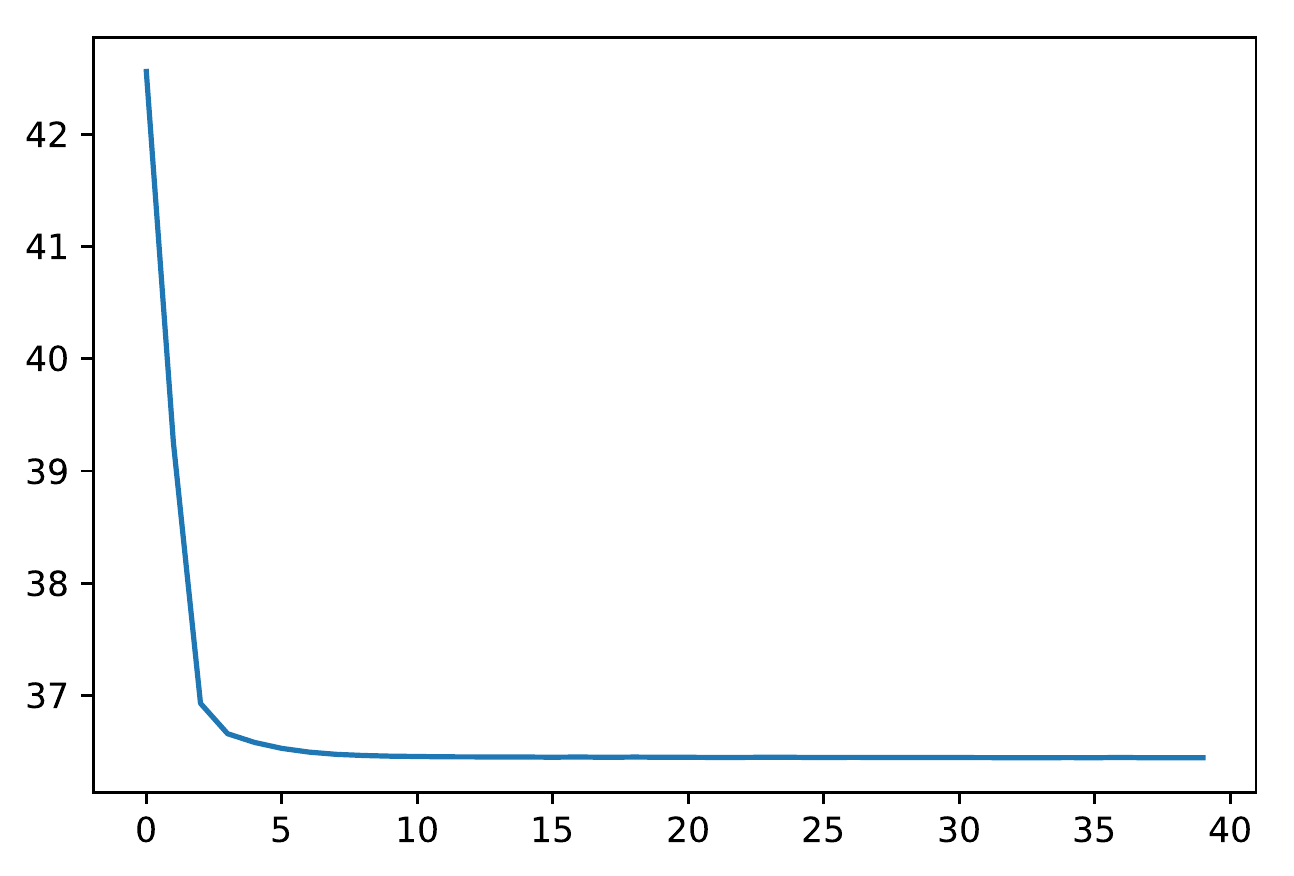}
\end{minipage}
\caption{Empirical squared distance without IS during training (left) and on the validation data (right).}
\label{fig1}
\end{figure}

It can be seen in Figure \ref{fig1} that the empirical squared distance on both, 
the training and validation data, decreases with a very similar decaying pattern. This provides a first validation of our 
approximation procedure. 

\begin{figure}[ht!] 
\centering
\begin{minipage}{.5\textwidth}
  \centering
  \includegraphics[width=.85\textwidth]{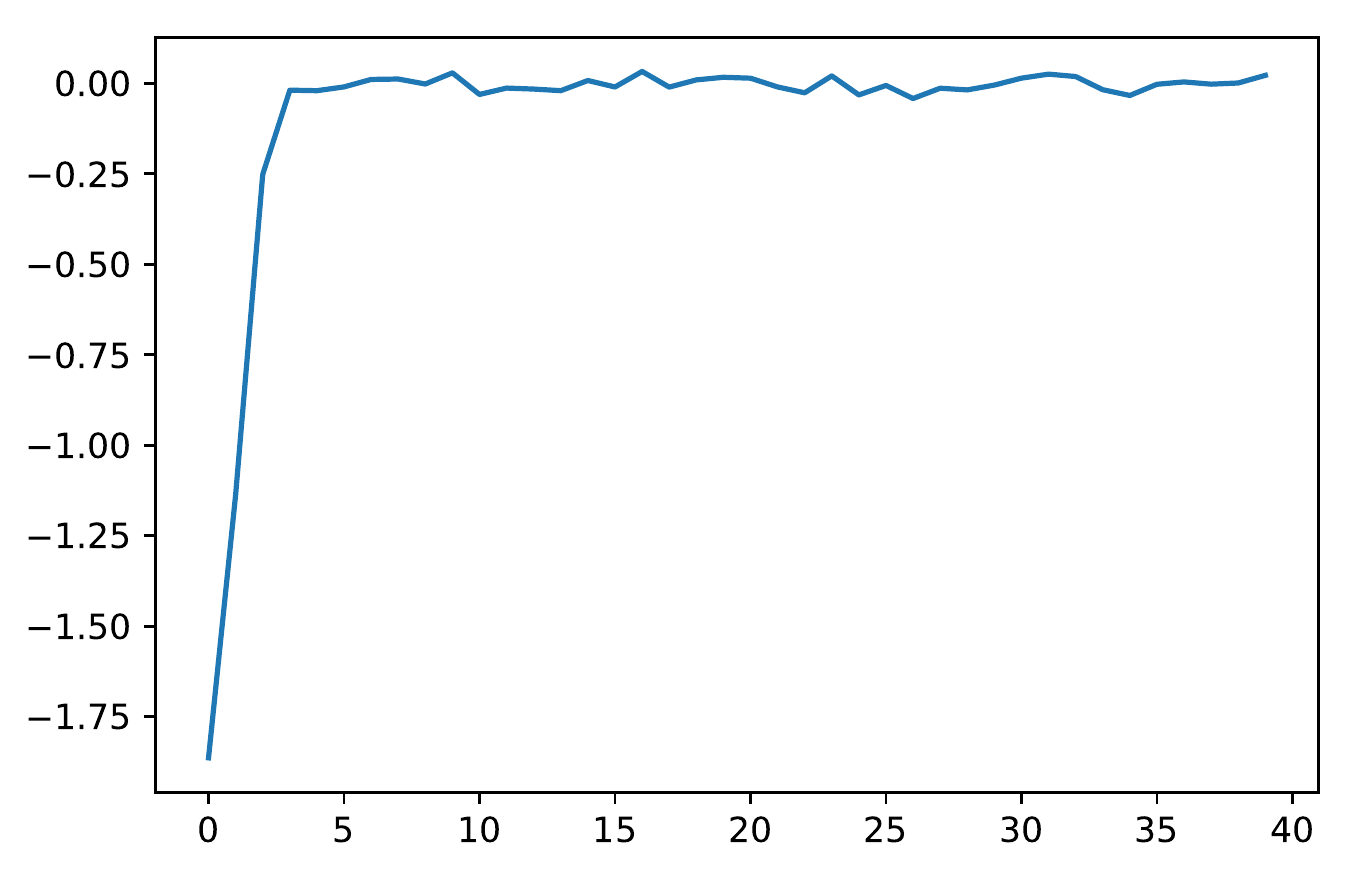}
\end{minipage}%
\begin{minipage}{.5\textwidth}
  \centering
  \includegraphics[width=.85\textwidth]{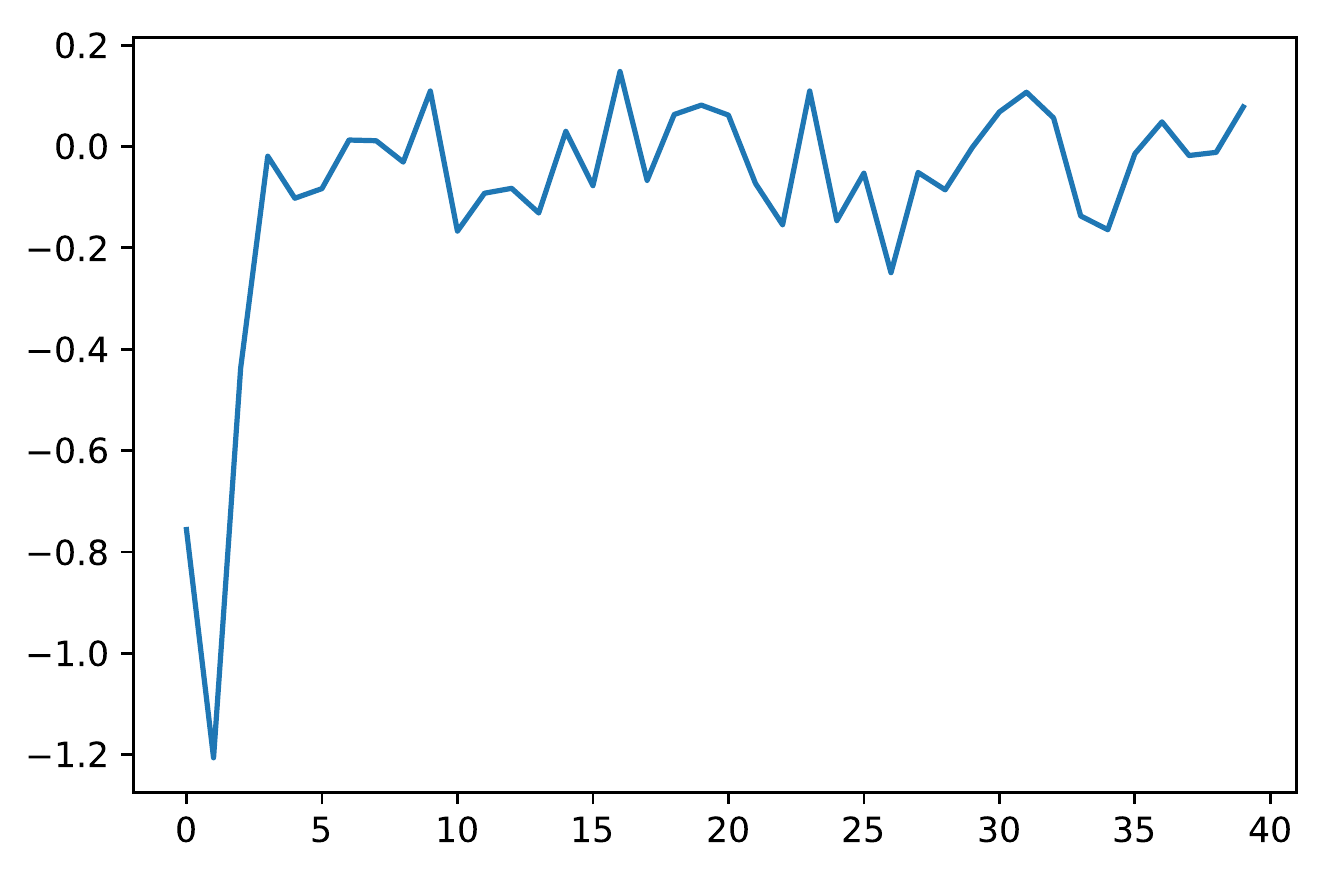}
\end{minipage}
\caption{Empirical evaluation of (a) (left) and (b) (right) of Section \ref{subsec:backtest}.}
\label{fig2}
\end{figure}

Figure \ref{fig2} shows the empirical evaluation of (a) and (b) of Section \ref{subsec:backtest}
on the test data after each training epoch.

\begin{figure}[ht!]
\centering
\begin{minipage}{.5\textwidth}
  \centering
  \includegraphics[width=.85\textwidth]{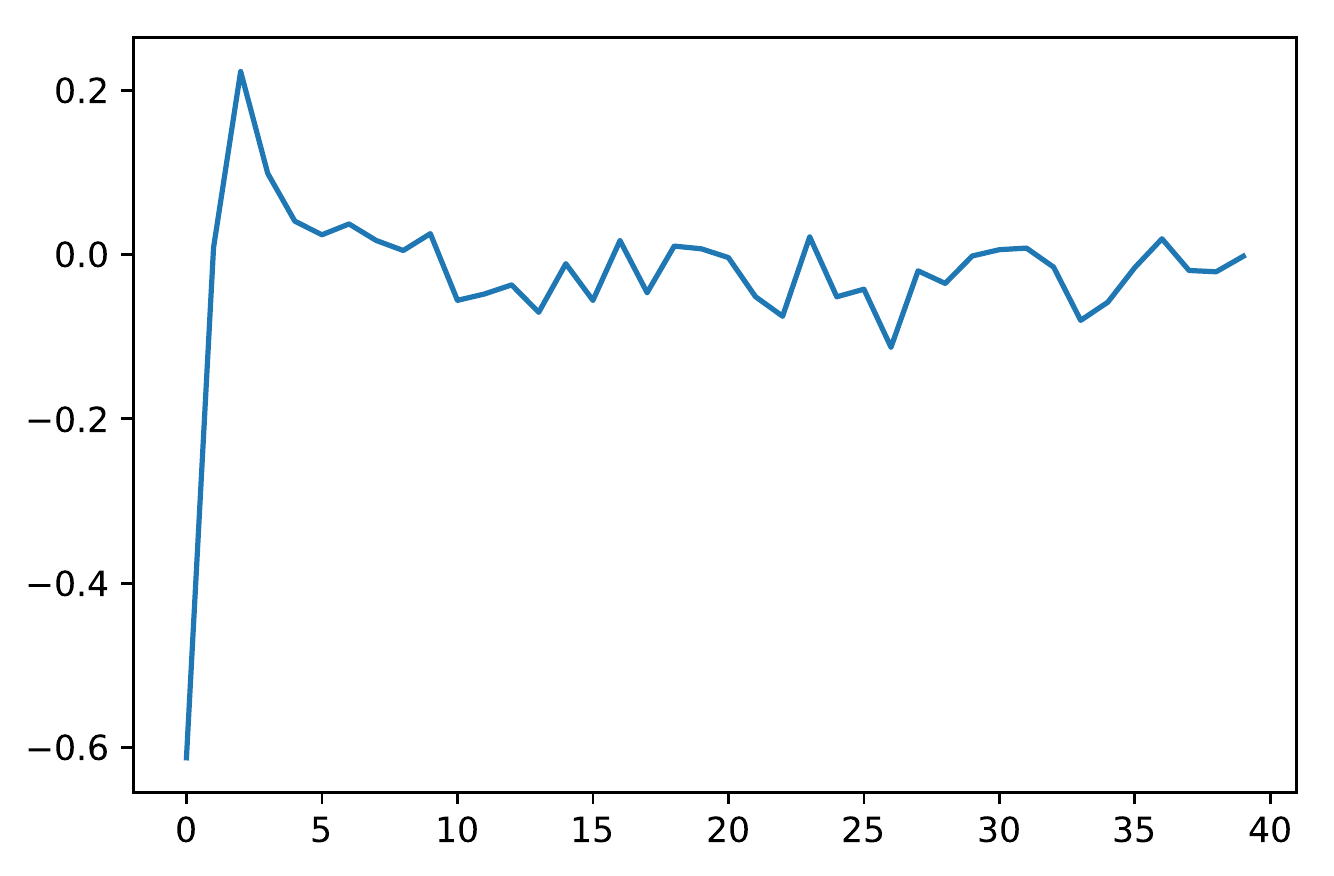}
\end{minipage}%
\begin{minipage}{.5\textwidth}
  \centering
  \includegraphics[width=.85\textwidth]{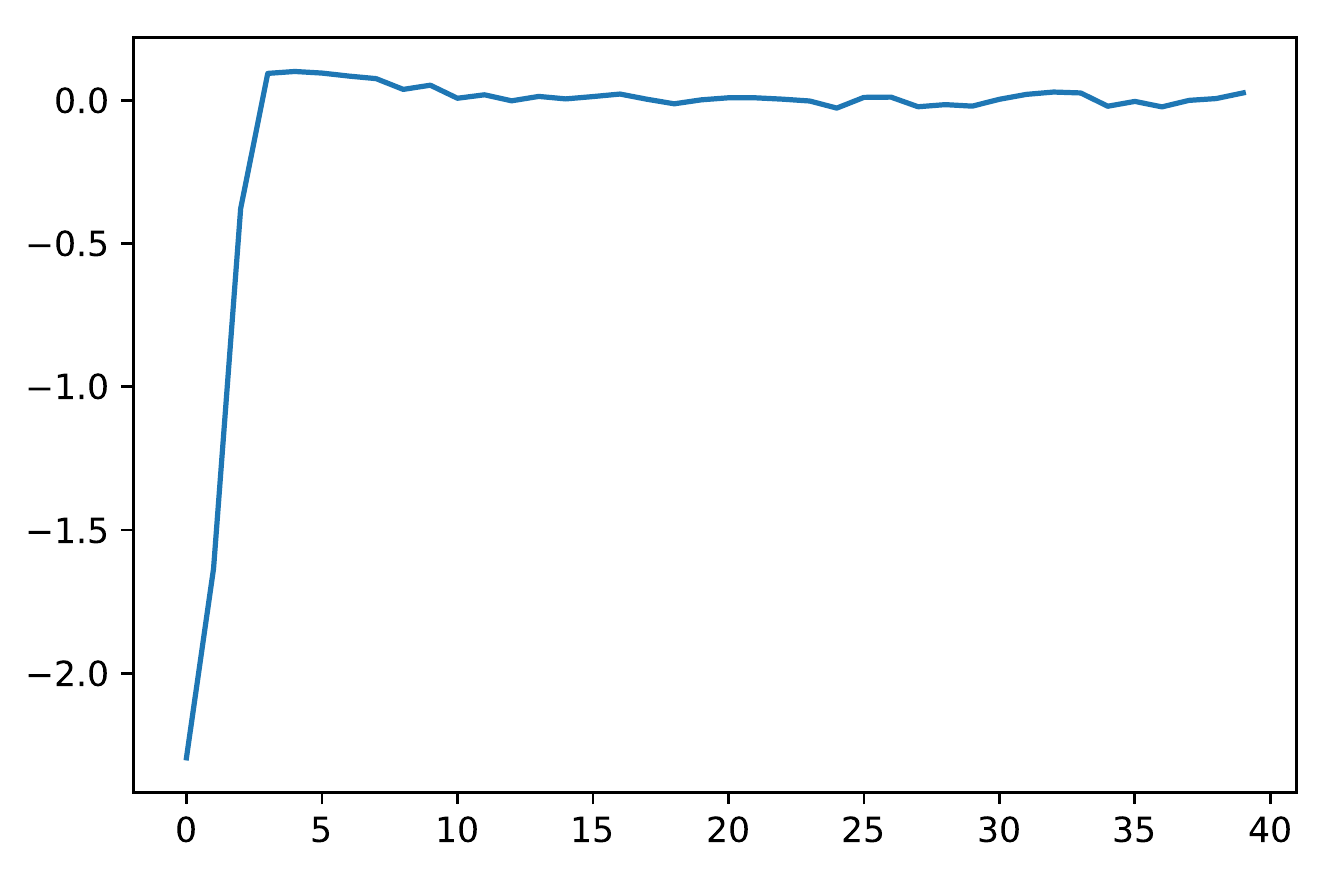}
\end{minipage}
\caption{Empirical evaluation of (c) of Section \ref{subsec:backtest} for the sets 
$B_1$ (left) and $B_2$ (right).} \label{fig3}
\end{figure}

Similarly, Figure \ref{fig3} illustrates the empirical evaluation of (c) of Section \ref{subsec:backtest} 
on the test data after each training epoch for the sets
\[ 
B_1 = \crl{x \in \R : x < s_{40\%}} \quad \mbox{and} \quad B_2 = \crl{x \in \R : x > s_{70\%}},
\]
where $s_{\beta}$ denotes the $\beta$-quantile of $X = S_{\tau}$.

Training, validation and testing with IS worked similarly.

\begin{figure}[ht!]
\centering
\begin{minipage}{.5\textwidth}
  \centering
  \includegraphics[width=.85\textwidth]{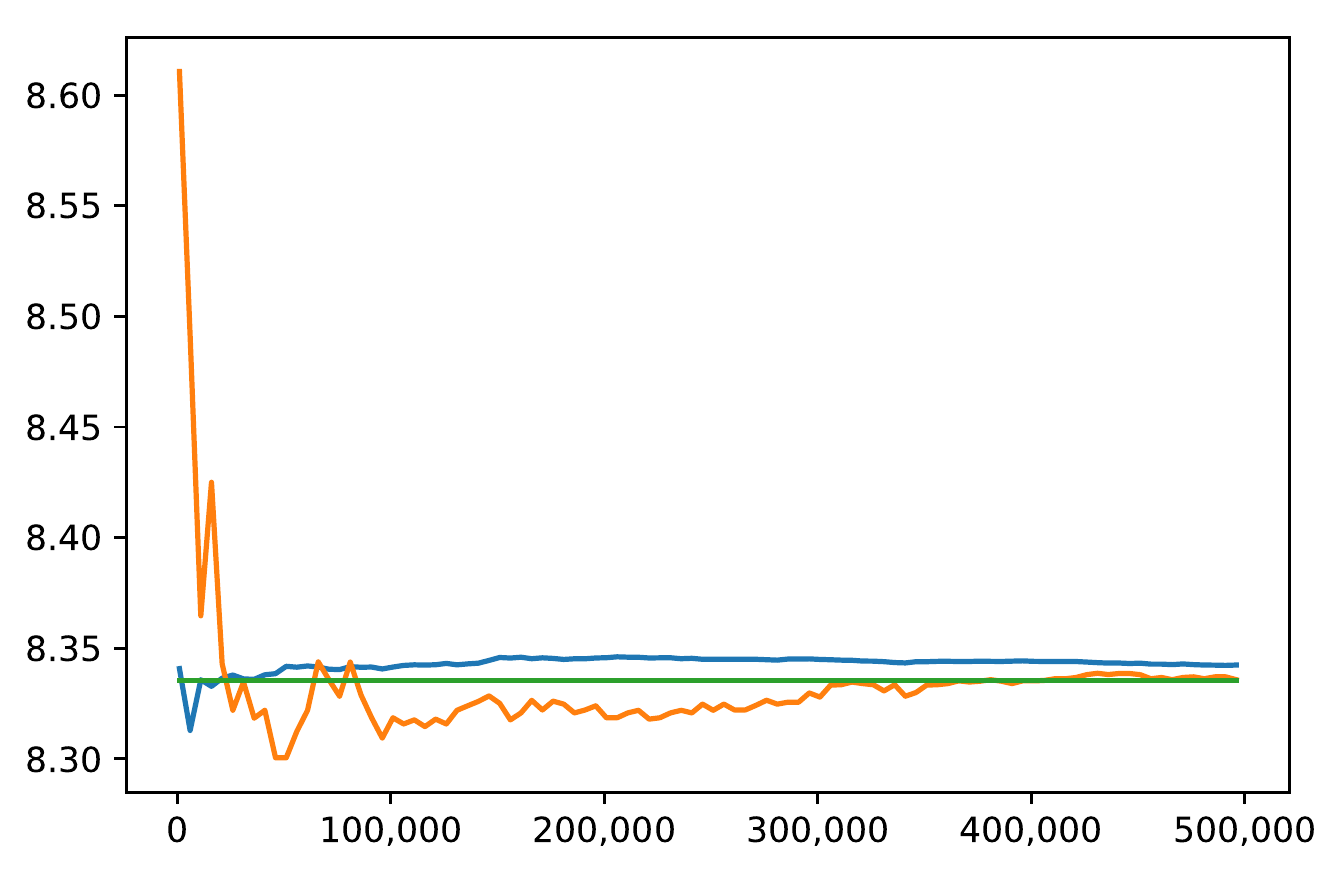}
  \end{minipage}%
\begin{minipage}{.5\textwidth}
  \centering
  \includegraphics[width=.85\textwidth]{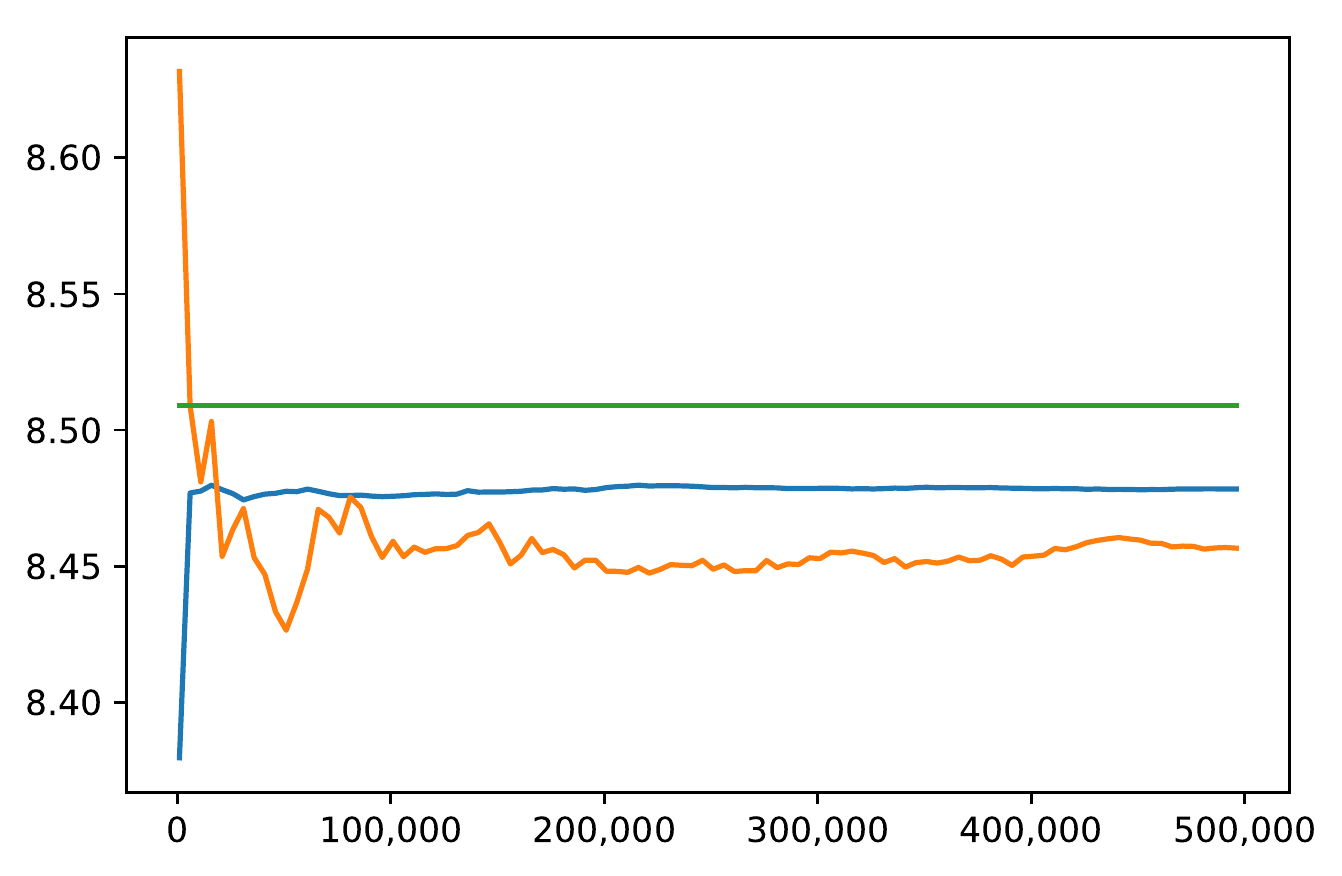}
  \end{minipage}
\caption{Convergence of the empirical 99.5\%-VaR (left) and 99\%-ES (right) without IS (orange) 
and with IS (blue) compared to the reference values obtained from the Black--Scholes formula (green).} \label{fig4}
\end{figure}

Once the network has been trained and tested, one can estimate VaR and ES numbers. Figure \ref{fig4} 
shows our results for increasing sample sizes. The left panel shows our estimate
of $\mbox{VaR}_{99.5\%}(L)$ without and with IS. Plugging the 0.5\%-quantile of $S_{\tau}$ into the 
Black--Scholes formula gives a reference value of 8.3356.
Our method yielded 8.3358 without and 8.3424 with importance sampling. 
The right panel shows our results for $\mbox{ES}_{99\%}(L)$. Transforming simulations of 
$S_{\tau}$ with the Black--Scholes formula and using the empirical ES estimate \eqref{PEst} 
resulted in a reference value of 8.509. Without importance sampling, the neural network learned a value 
of 8.456 versus 8.478 with importance sampling. It can be seen that in both cases, IS made the method 
more efficient. It has to be noted that for increasing sample sizes, the VaR and ES estimates converge to 
the corresponding risk figures in the neural network model, which are not exactly equal to their analogs in 
the correct model. But it can be seen that the blue lines are close to their final values after very few simulations.

\subsection{Portfolio of call and put options} 
\label{Subsec:20opt}

In our second example we introduce a larger set of risk factors. We consider a portfolio of 20 short
call and put options on different underlying assets with initial prices $s^i_0 > 0$ and dynamics
\[
dS^i_t = \mu_i S^i_t dt + \sigma_i S^i_t dW^{\p,i}_t = r S^i_t dt + \sigma_i S^i_t dW^{\q,i}_t
\]
for $\p$-Brownian motions $W^{\p,i}$ and $\q$-Brownian motions $W^{\q,i}$, $i =1, \dots, 20$
such that $(W^{\p,1}, \dots, W^{\p,20})$ is a multivariate Gaussian process under $\p$ with an
instantaneous correlation of 30\% between different components and 
$(W^{\q,1}, \dots, W^{\q,20})$ is a multivariate Gaussian process under $\q$, also with
instantaneous correlation of 30\% between different components.
We set $s^i_0 = 100$ for all $i = 1, \dots, 20$ and $r = 1\%$. The drift and volatility parameters 
are assumed to be $\mu_i = \mu_{10 + i} = (2.5 + i/2) \%$ and $\sigma_i = \sigma_{10+i} 
= (14+i) \%$, $ i=1,\ldots,10$.
As in the first example, we choose a maturity of $T = 1/3$ and a risk horizon of $\tau = 1/52$. 
We assume all options have the same strike price $K = 100$. Then the time-$\tau$ value of the liability is 
\[
L = e^{-r(T-\tau)} \E \edg{\sum_{i=1}^{10} (S^i_T -K)^+ 
+ \sum_{i=11}^{20} (K-S^i_T)^+ \Big| \, X},
\]
where $X$ is the vector $\left(S_{\tau}^1,\ldots,S_{\tau}^{20}\right)$.

\begin{figure}[ht!]
\centering
\begin{minipage}{.5\textwidth}
  \centering
  \includegraphics[width=.85\textwidth]{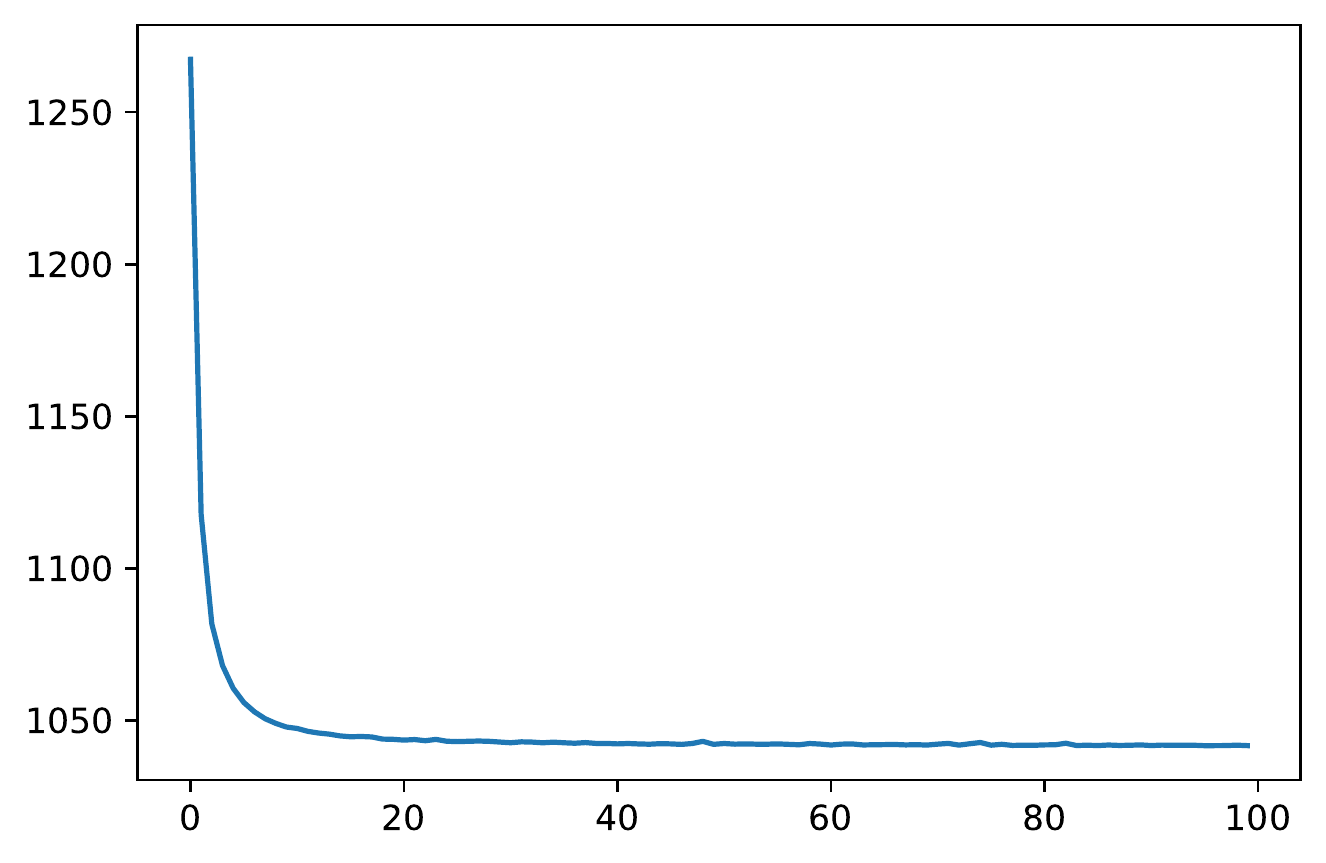}
\end{minipage}%
\begin{minipage}{.5\textwidth}
  \centering
  \includegraphics[width=.85\textwidth]{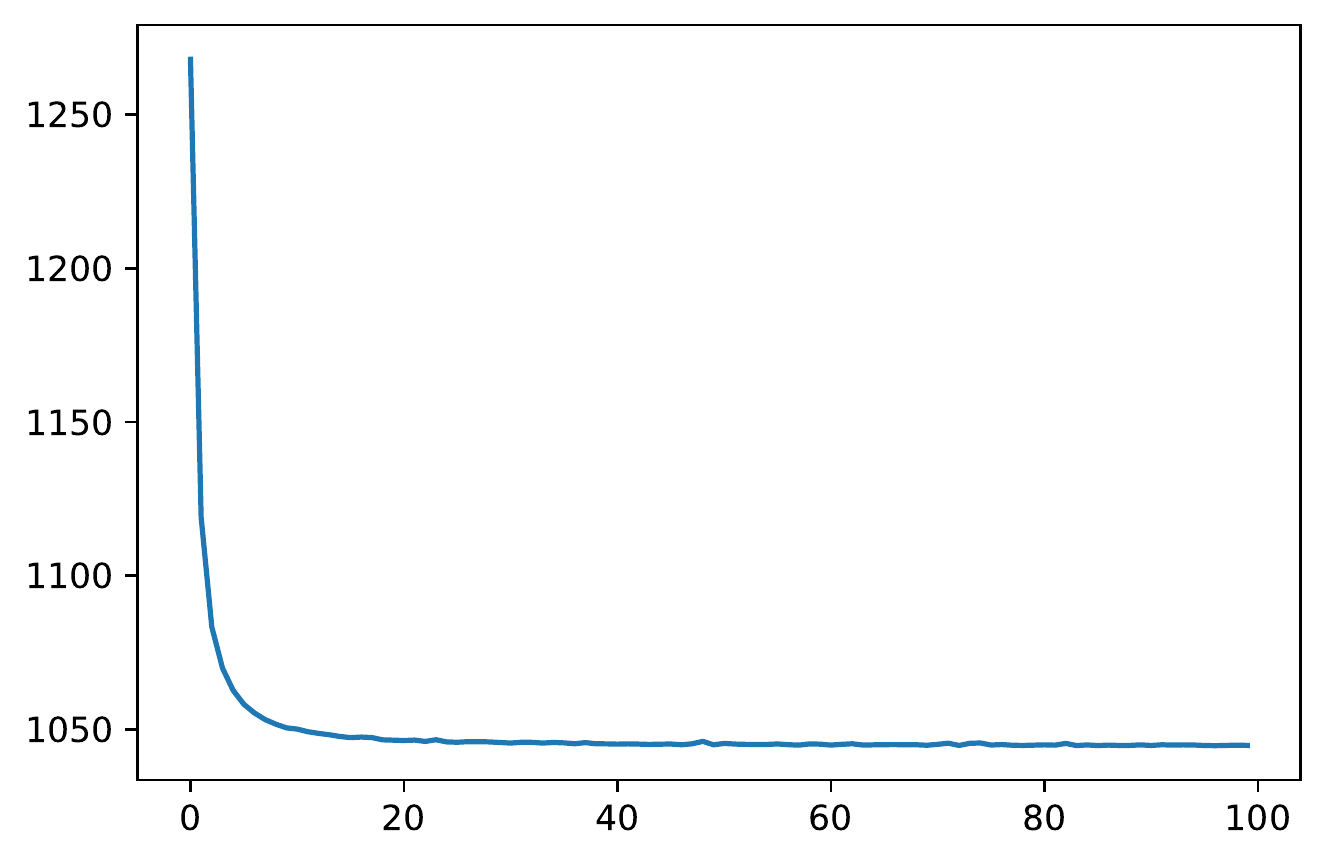}
\end{minipage}
\caption{Empirical squared distance without IS during training (left) and on the validation data (right).} \label{fig:multi}
\end{figure}

In this example we trained a neural network with two hidden layers containing 15 nodes each.
We first simulated according to $\p \otimes \q$ and trained for 100 epochs. Figure \ref{fig:multi} shows 
the decay of the empirical squared distance on the training and validation data set. 

\begin{figure}[ht!]
\centering
\begin{minipage}{.5\textwidth}
  \centering
  \includegraphics[width=.85\textwidth]{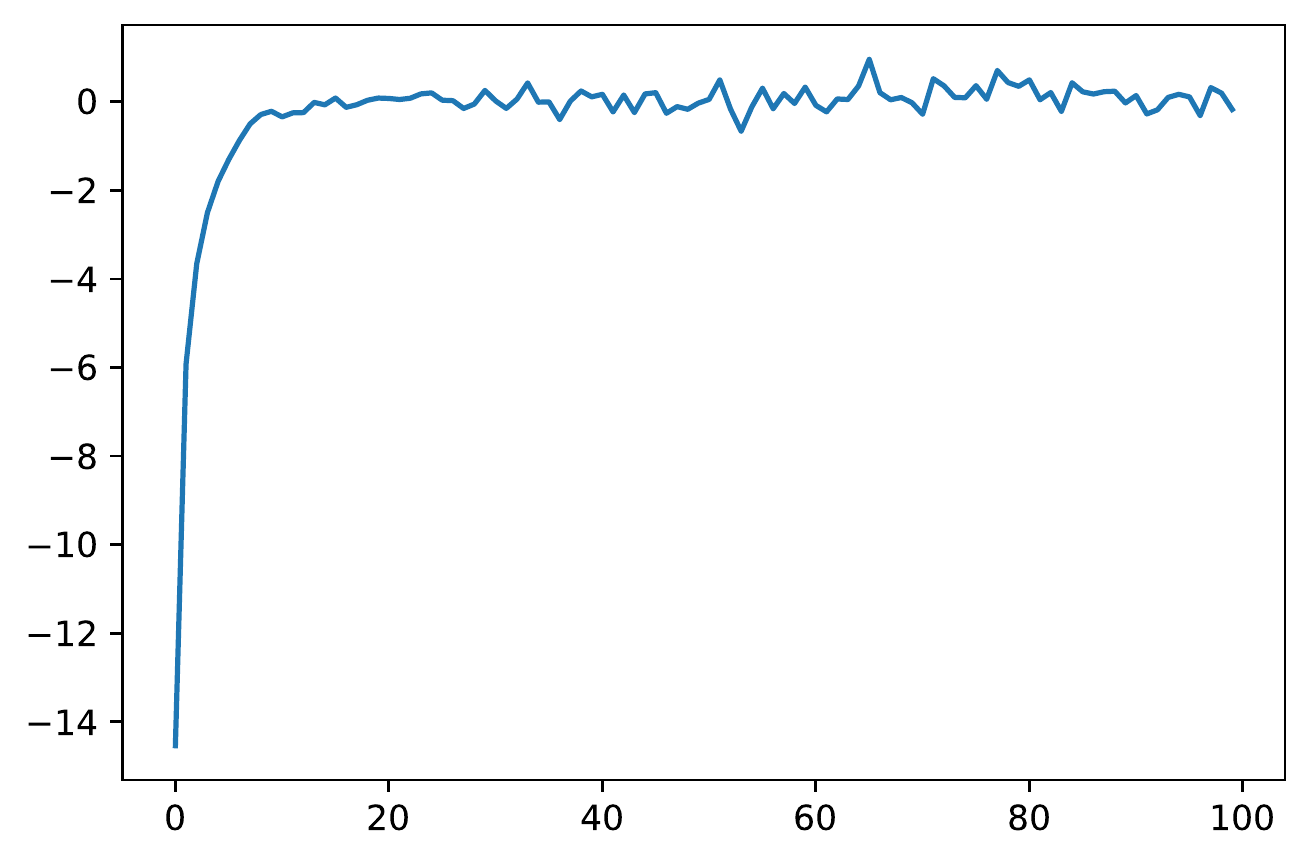}
\end{minipage}%
\begin{minipage}{.5\textwidth}
  \centering
  \includegraphics[width=.85\textwidth]{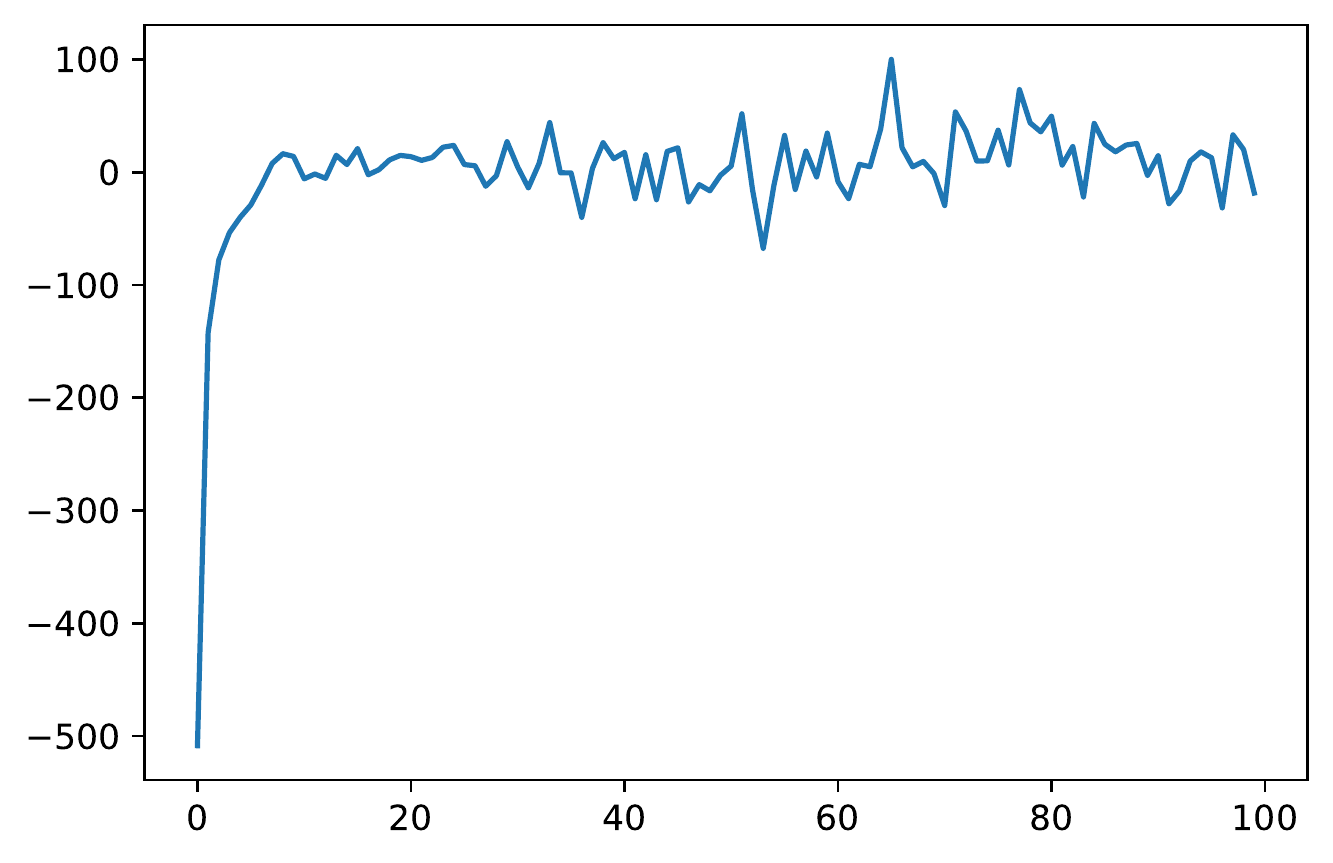}
 \end{minipage}
\caption{Empirical evaluation of (a) (left) and (b) (right) of Section \ref{subsec:backtest} 
under $\p^{\nu} \otimes \q$ corresponding to $\alpha = 99\%$.}
\label{fig6}
\end{figure}

After training and testing the network under $\p \otimes \q$, we did the same 
under $\p^{\nu} \otimes \q$ for the IS distributions $\nu$ resulting from the 
procedure of Section \ref{Subsec:nis} for $\alpha = 99.5\%$ (for VaR) and $99\%$ (for ES).
The two plots in Figure \ref{fig6} show the empirical evaluations of (a) and (b) on the test 
data under the IS measure $\p^{\nu} \otimes \q$ corresponding to $\alpha = 99\%$.

\begin{figure}[ht!]
\centering
\begin{minipage}{.5\textwidth}
  \centering
  \includegraphics[width=.85\textwidth]{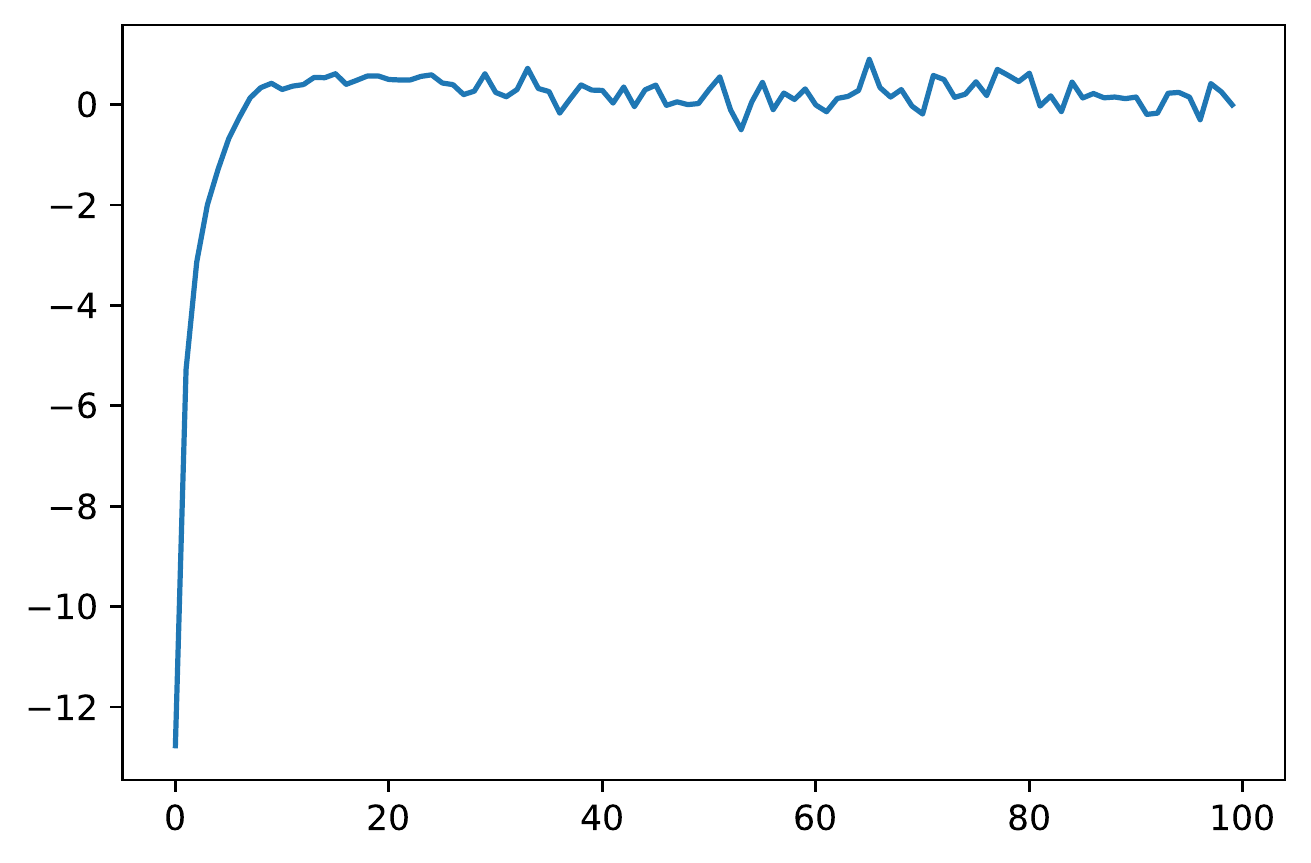}
\end{minipage}%
\begin{minipage}{.5\textwidth}
  \centering
  \includegraphics[width=.85\textwidth]{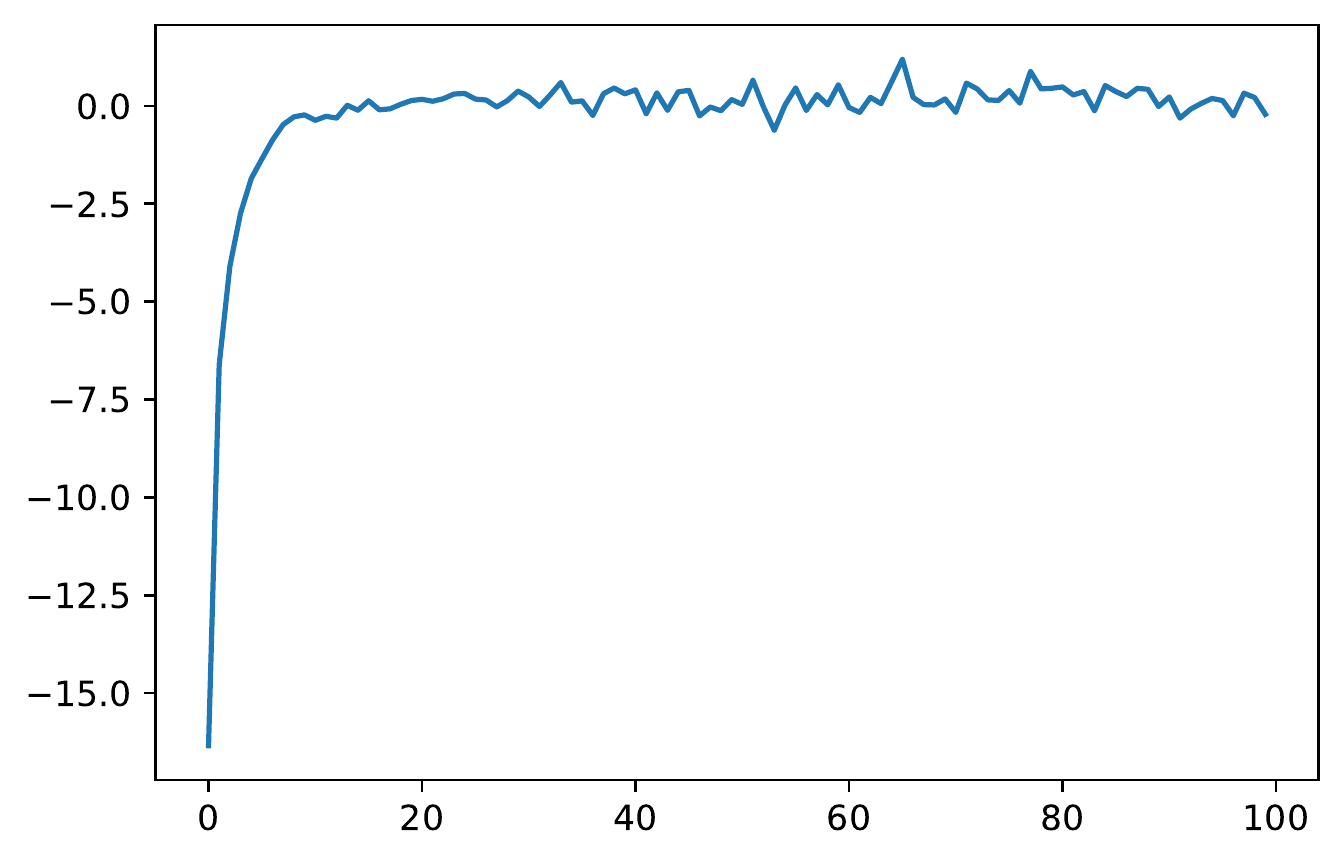}
  \end{minipage}
\caption{Empirical evaluation of (c) of Section \ref{subsec:backtest} for the sets 
$B_1$ (left) and $B_2$ (right) under $\p^{\nu} \otimes \q$ corresponding to $\alpha = 99\%$.}
\label{fig7}
\end{figure}

As an additional test, we consider the two sets 
\[
B_1=\{x \in \R^{20} : x_i > s^i_{20\%} \mbox{ and } x_{10+i} < s^{10+i}_{80\%} \mbox{ for }  i =1 ,2,3 \}
\]
and
\[
B_2 =\{x \in \R^{20} : x_i < s^i_{80\%} \mbox{ and } x_{10+i} > s^{10+i}_{20\%} \mbox{ for }  i =1 ,2,3 \},
\]
where $s^i_{\beta}$ is the $\beta$-quantile of $S^i_{\tau}$ under $\p^{\nu} \otimes \q$, 
and evaluate (c) on the test data generated under the IS distribution $\p^{\nu} \otimes \q$
corresponding to $\alpha = 99\%$. The results are depicted in Figure \ref{fig7}.

\begin{figure}[ht!]
\centering
\begin{minipage}{.5\textwidth}
  \centering
  \includegraphics[width=.85\textwidth]{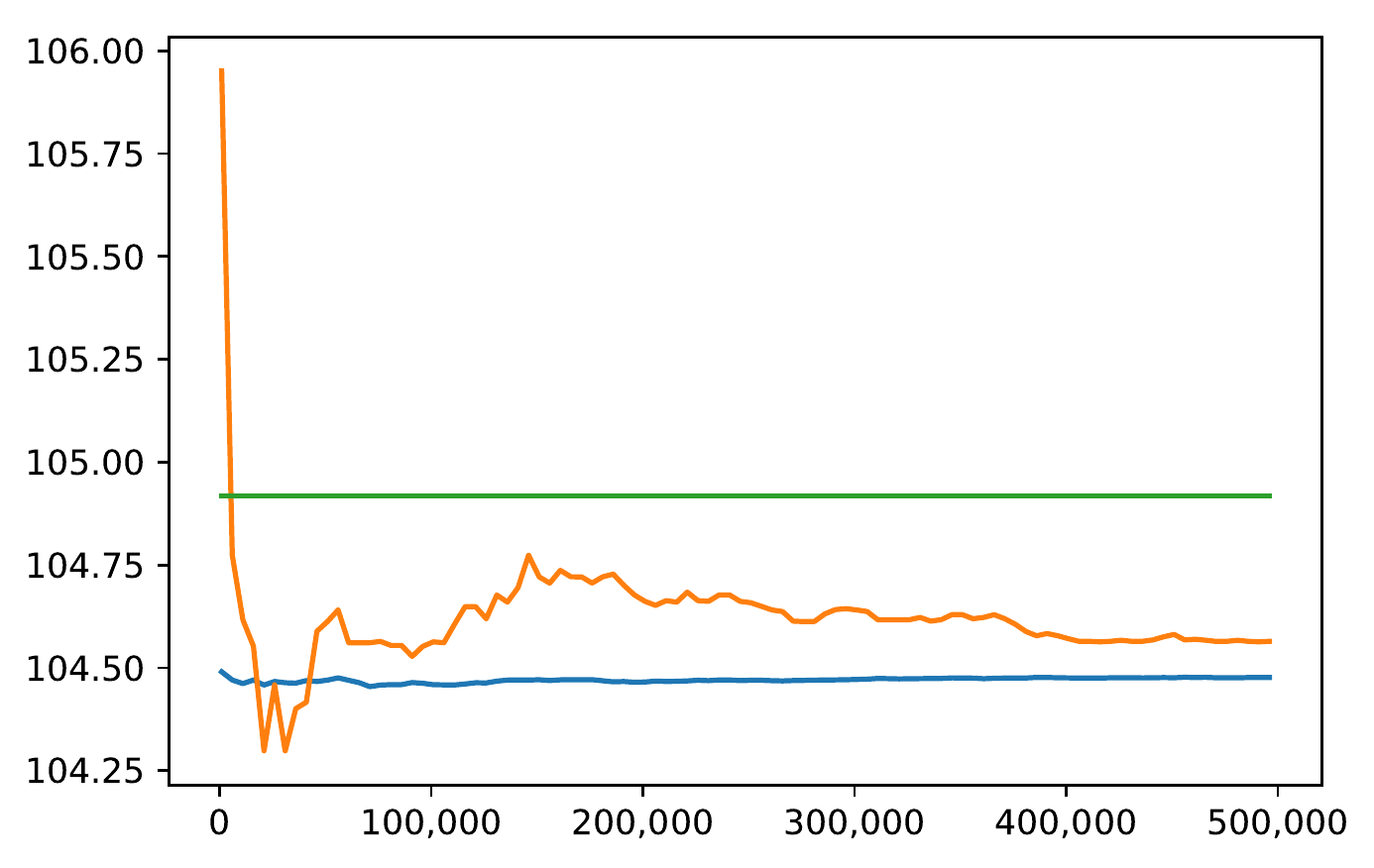}
\end{minipage}%
\begin{minipage}{.5\textwidth}
  \centering
  \includegraphics[width=.85\textwidth]{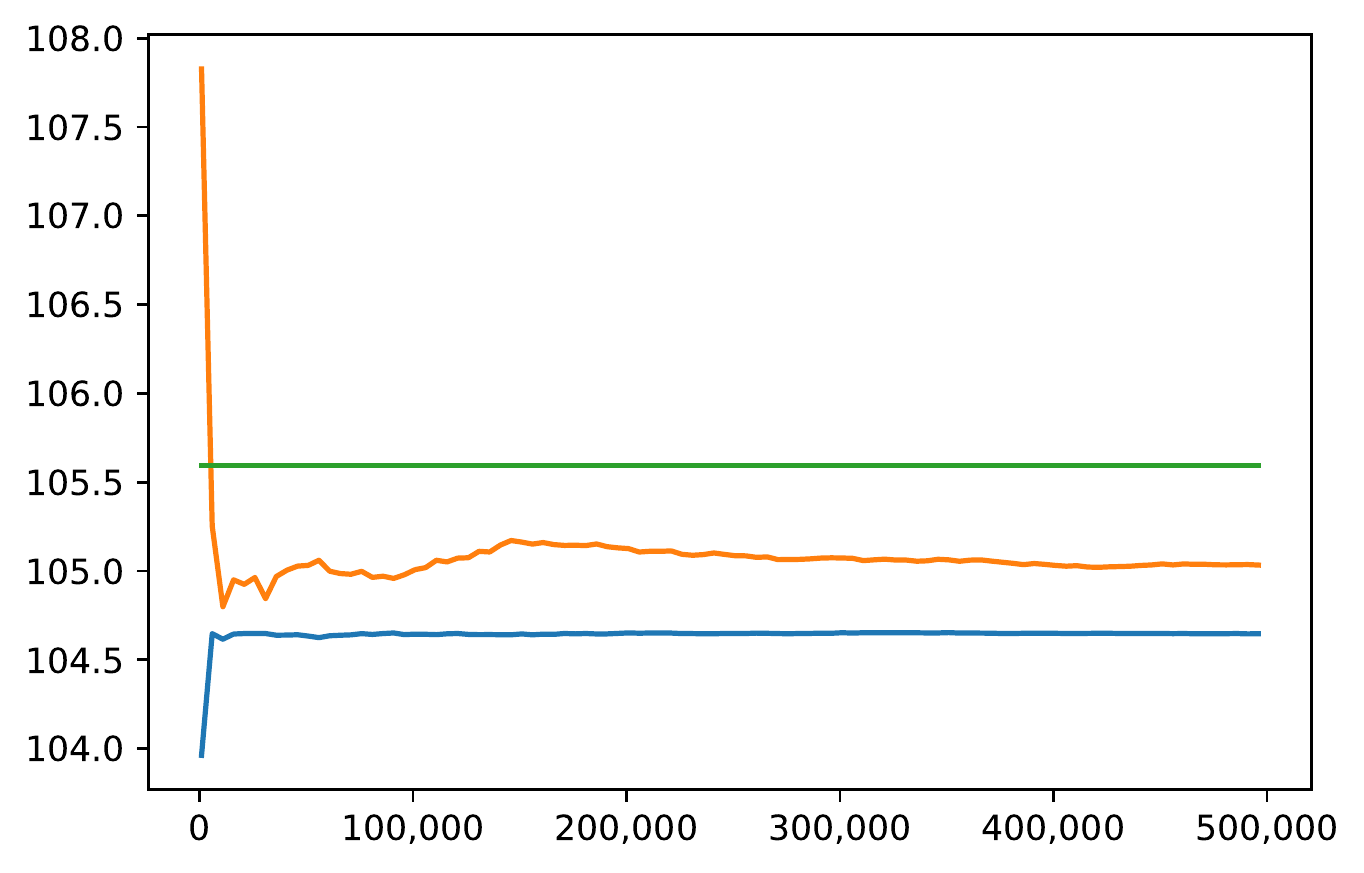}
\end{minipage}
\caption{Convergence of the empirical 99.5\%-VaR (left) and 99\%-ES (right) without IS (orange) and
with IS (blue) compared to the reference values obtained from the Black--Scholes formula (green).} \label{fig8}
\end{figure}

After training and testing, we generated simulations of $X$ to estimate $\mbox{VaR}_{99.5\%}(L)$ 
and $\mbox{ES}_{99\%}(L)$. The convergence for increasing sample sizes is shown in Figure \ref{fig8}.
The reference values, 104.92 for $99.5\%$-VaR and 105.59 for $99\%$-ES, were obtained from the
empirical estimates \eqref{PEst} by simulating $S_{\tau}$ and using the Black--Scholes formula for each 
of the 20 options. The neural network estimates of $99.5\%$-VaR were 104.56 without and 104.48 with IS.
Those of $99\%$-ES were 105.03 without and 104.65 with IS.
In both cases, IS made the procedure more efficient. 

\subsection{Variable annuity with GMIB} 
\label{Subsec:va}

As a third example we study a variable annuity (VA) with guaranteed minimum income benefit (GMIB).
We consider the contract analyzed by \cite{bauer} using polynomial regression. 
At time $0$ the contract is sold to an $x$-year old policyholder. If she is still alive at maturity $T$,
she can choose between the balance of an underlying investment account and a lifetime 
annuity. Therefore, in case of survival, the time-$T$ value of the contract is 
\[
\max \crl{S_T \, ,\, b \, a_{x+T}(T)},
\]
where $S_T$ is the account value, $b$ a guaranteed rate and $a_{x+T}(T)$ the time-$T$ value 
of an annuity paying one unit of currency to a $(x+T)$-year old policyholder 
at times $T+1, T+2, ...$ for as long as the person lives. 

The contract is exposed to three types of risk: investment risk, interest rate risk and mortality risk. 
We suppose the log-account value $q_t = \log(S_t)$, the interest rate $r_t$ and the mortality 
rate $\mu_{x+t}$ of our policyholder start from known constants $q_0, r_0, \mu_x$ and 
for $x + t \le 120$, have $\p$-dynamics
\begin{align*}
dq_t&=\left(m-\frac{1}{2}\sigma_S^2\right)dt+\sigma_SdW_t^{\p,S},\\
dr_t&=\zeta(\gamma-r_t)dt+\sigma_rdW_t^{\p,r},\\
d\mu_{x+t}& =\kappa\mu_{x+t}dt+\sigma_{\mu} dW_t^{\p, \mu},
\end{align*}
for given parameters $m, \zeta,\gamma,\kappa, \sigma_S,\sigma_r, \sigma_{\mu}$ 
and $\p$-Brownian motions $W^{\p,S},$ $W^{\p,r}$ and $W^{\p,\mu}$ forming a 
three-dimensional Gaussian process with instantaneous correlations $\rho_{12}$, $\rho_{13}$ and $\rho_{23}$. 
We assume that our policyholder does not live longer than 120 years. Therefore, we set 
$\mu_{x+t} \equiv \infty$ for $x+ t > 120$. The dynamics for $x+ t \le 120$ under the 
risk-neutral probability $\q$ are assumed to be 
\begin{align*}
dq_t&=\left(r_t-\frac{1}{2}\sigma_S^2\right)dt+\sigma_SdW_t^{\q,S},\\
dr_t&=\zeta(\bar{\gamma}-r_t)dt+\sigma_rdW_t^{\q,r},\\
d\mu_{x+t}& =\kappa\mu_{x+t}dt+\sigma_{\mu} dW_t^{\q,\mu},
\end{align*}
where $W^{\q,S},$ $W^{\q,r},$ $W^{\q, \mu}$ are $\q$-Brownian motions 
constituting a three-dimensional Gaussian process
with the same instantaneous correlations as the corresponding $\p$-Brownian motions. 
As \cite{bauer}, we assume there is no risk premium for mortality and a constant risk 
premium $\lambda$ for interest rate risk, such that $\bar{\gamma}= \gamma-\lambda\sigma_r/\zeta.$
Provided that the policyholder is still alive at the risk horizon $\tau < T$, the value of the
contract at that time is 
\[
L = \E^{\q}\edg{e^{-\int_{\tau}^T r_s + \mu_{x+s} ds} \max\left\{e^{q_T},b \, a_{x+T}(T) \right\}\mid X},
\]
where we denote $X = (q_{\tau}, r_{\tau}, \mu_{x + \tau})$. Discounting with $r_s + \mu_{x+s}$ takes into 
account that the policyholder might die between $\tau$ and $T$. On the other hand, a possible death 
time between $0$ and $\tau$ is not considered. This results in a conservative estimate 
of the capital requirement for the issuer of the contract. Alternatively, one could model the loss
as $I_A L$, where $A \subseteq \Omega$ is the event that the policyholder survives until time $\tau$.

\begin{figure}[ht!]
\centering
\begin{minipage}{.5\textwidth}
  \centering
  \includegraphics[width=.85\textwidth]{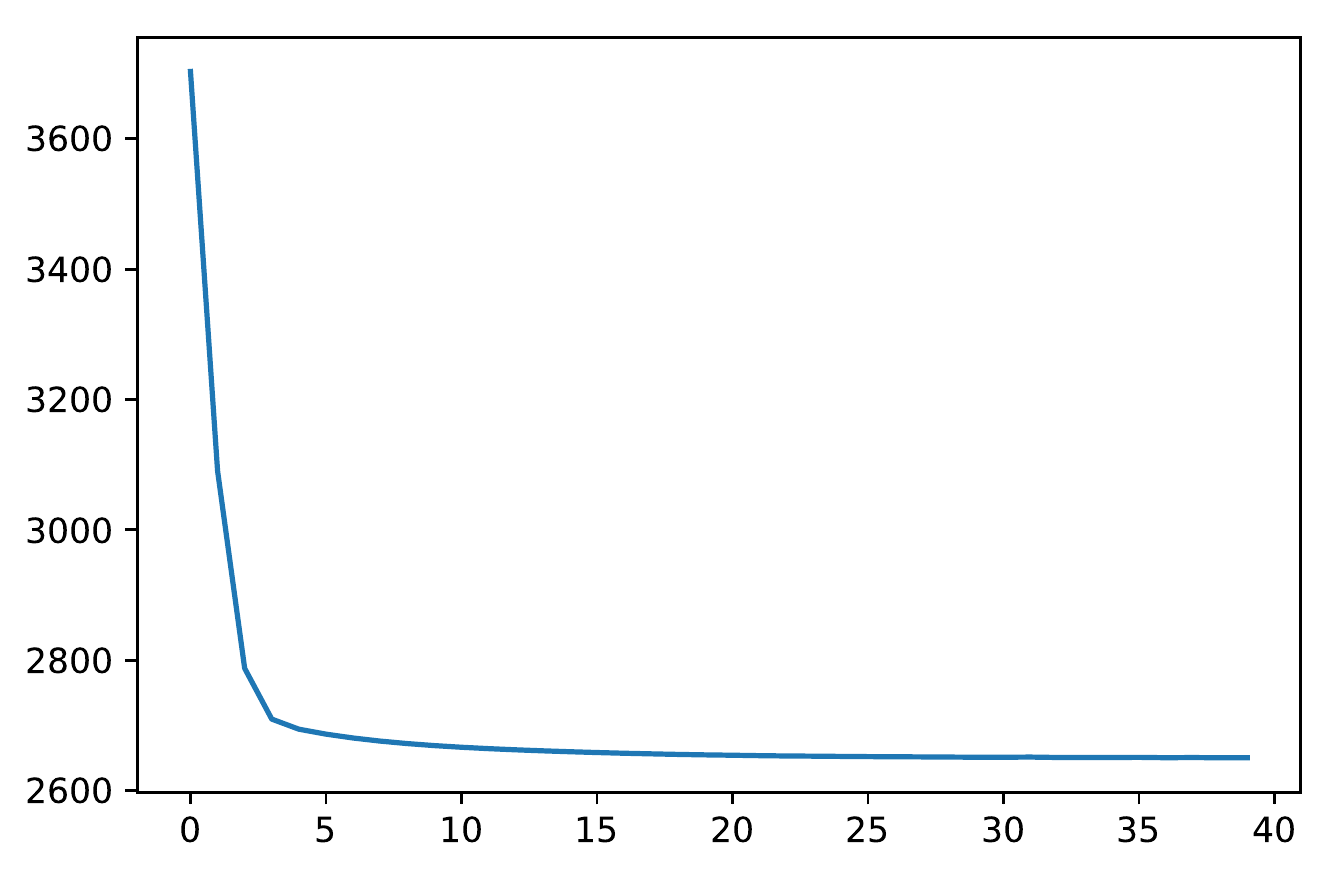}
\end{minipage}%
\begin{minipage}{.5\textwidth}
  \centering
  \includegraphics[width=.85\textwidth]{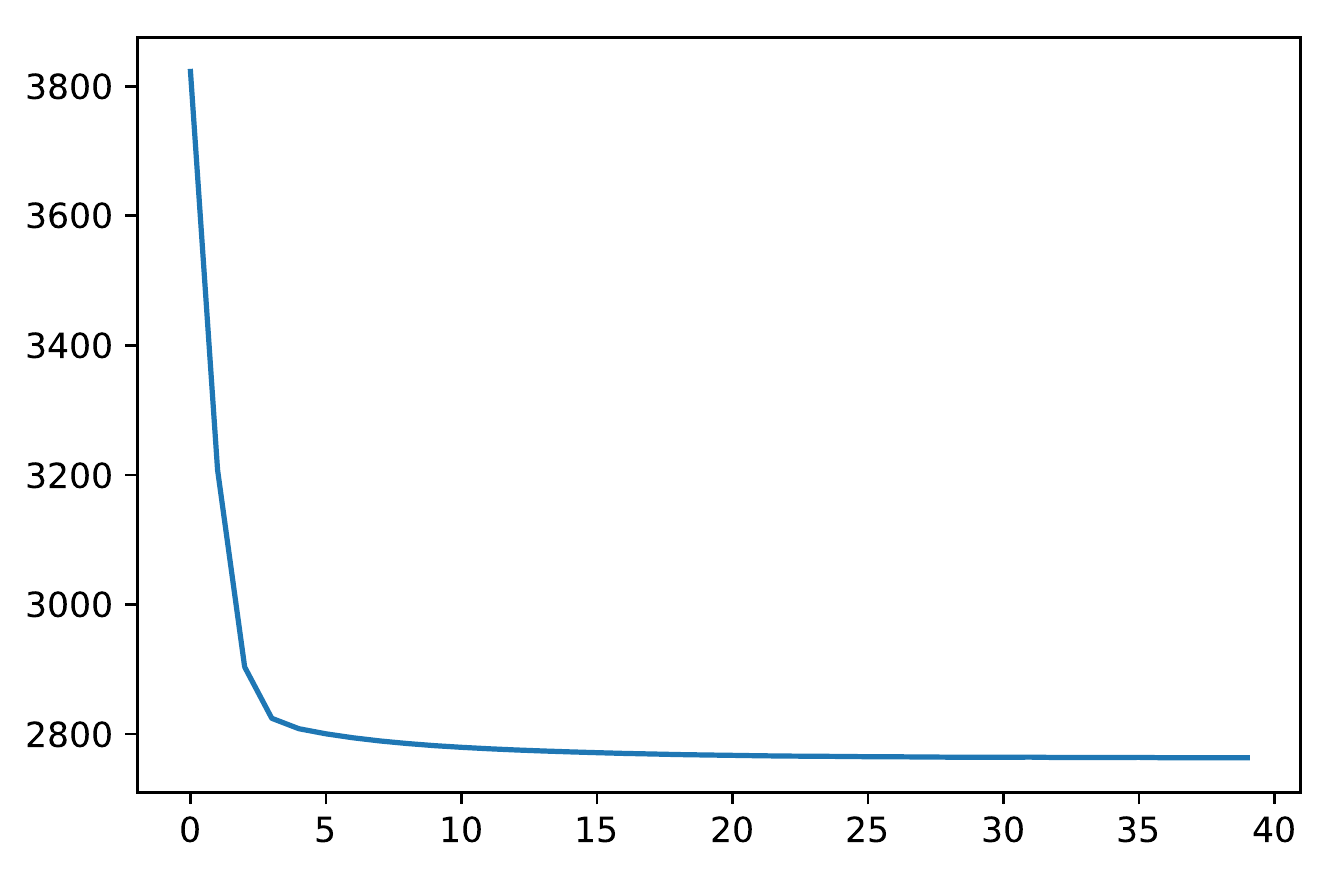}
\end{minipage}
\caption{Empirical squared distance without IS during training (left) and on the validation data (right).}
\label{fig:VA3}
\end{figure}

We follow \cite{bauer} and set $x = 55$, $\tau = 1$, $T = 15$, $b = 10.792$, $q_0=4.605$, $m = 5\%$, 
$\sigma_S=18\%$, $r_0= 2.5\%$, $\zeta= 25\%$, $\gamma= 2\%$, $\sigma_r=1\%$, $\lambda= 2\%$,
$\mu_x= 1\%$, $\kappa= 7\%$, $\sigma_{\mu}=0.12\%$, $\rho_{12}=-30\%$, $\rho_{13}=6\%$, $\rho_{23}=-4\%$. 
Then 
\[
a_{x+T}(T) = \sum_{k = 1}^{50} {}_{k}E_{x+T}(T),
\]
where 
\[
{}_{k}E_{x+t}(t) = \E^{\q} \edg{e^{- \int_t^{t+k} r_s + \mu_{x +s} ds} \mid r_t, \mu_{x+t}}
\]
is the time-$t$ value of a pure endowment contract with maturity $t+k$. Since $r$ and $\mu$ are affine, one has
\[
{}_{k}E_{x+t}(t) = F(t, k, r_t, \mu_{x+t}) 
\]
for the function $F(t,k,r_t,\mu_{x+t})=A(t,t+k)e^{-B_r(t,t+k)r_t-B_\mu(t,t+k)\mu_{x+t}}$, with $A, B_r,$ and $B_\mu$ 
as given in the Appendix of \cite{bauer}.
Moreover, one can write 
\[
L = {}_{T-\tau} E_{x + \tau}(\tau) \E^{\q_E} \edg{\max\left\{e^{q_T},b \, a_{x+T}(T)\right\}\mid X}
\]
for the probability measure $\q_E$ given by 
\[
\frac{d\q_E}{d\q} = \frac{\exp\brak{- \int_0^T r_s + \mu_{x+s} ds}}{\E^{\q} \exp\brak{- \int_0^T r_s + \mu_{x+s} ds}}.
\]
Under $\p$, $X = (q_{\tau}, r_{\tau}, \mu_{\tau})$ is a three-dimensional normal vector,
and the conditional $\q_E$-distribution of $(q_T, r_T, \mu_{x+T})$ given $X$ is normal too (the precise form 
of these distributions is given in the Appendix of \cite{bauer}).
This makes it possible to efficiently simulate $(X,Y)$ under $\p \otimes \q_E$, where
\[
Y = F(\tau, T - \tau, r_{\tau}, \mu_{x + \tau}) \max \crl{e^{q_T},b \sum_{k=1}^{50} F(T,k, r_T, \mu_{x+T})}.
\]
To approximate $L$ we chose a network with two hidden layers containing 4 nodes each and trained it 
for 40 epochs. Figure \ref{fig:VA3} shows the empirical squared distance 
without IS on the training and validation data set.

\begin{figure}[ht!]
\centering
\begin{minipage}{.5\textwidth}
  \centering
  \includegraphics[width=.85\textwidth]{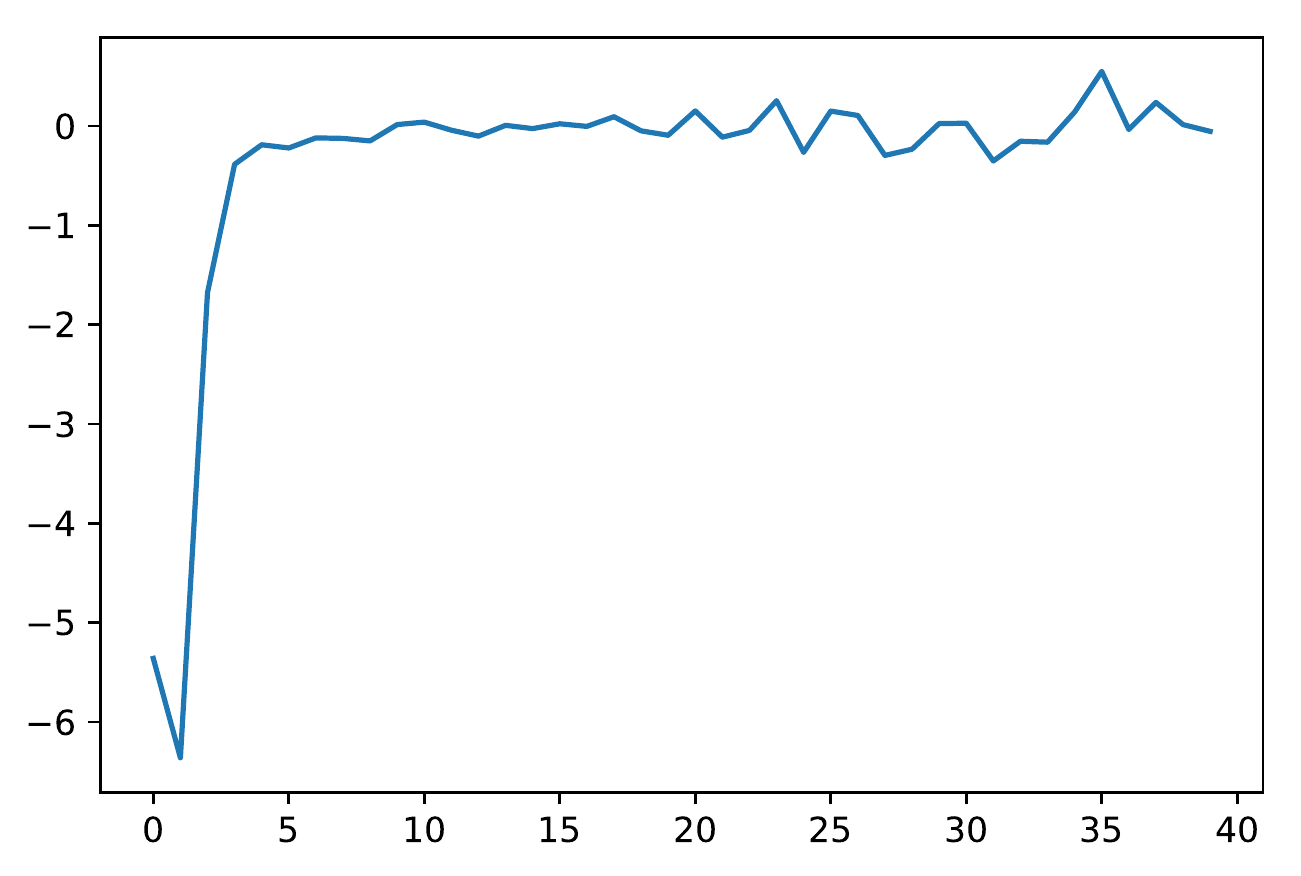}
\end{minipage}%
\begin{minipage}{.5\textwidth}
  \centering
  \includegraphics[width=.85\textwidth]{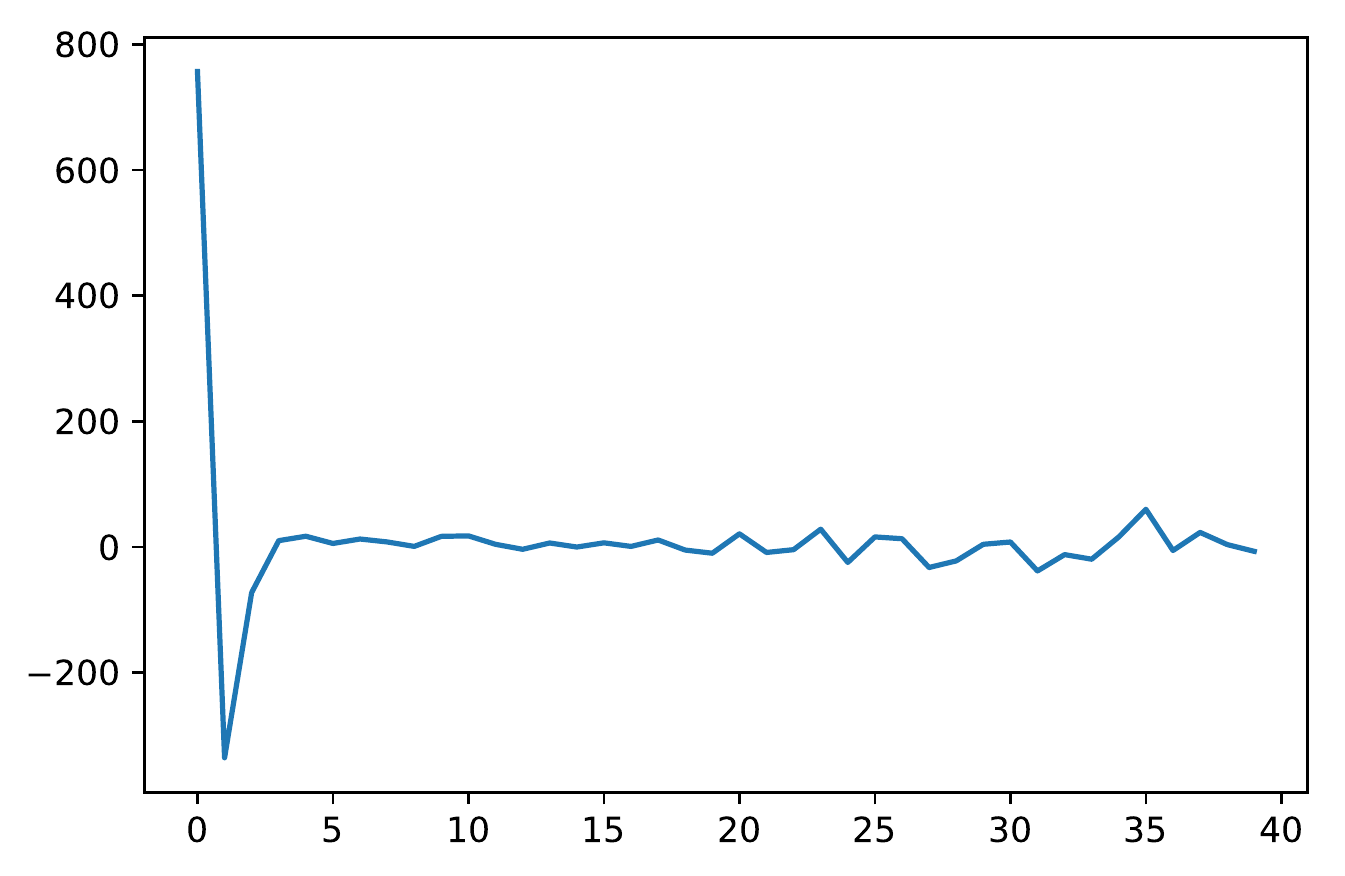}
\end{minipage}
\caption{Empirical evaluation of (a) (left) and (b) (right) of Section \ref{subsec:backtest} 
under $\p \otimes \q_E$.} \label{VAatest}
\end{figure}

The panels in Figure \ref{VAatest} illustrate the empirical evaluations of the test criteria (a) and (b) 
from Section \ref{subsec:backtest}.

\begin{figure}[ht!]
\centering
\begin{minipage}{.5\textwidth}
  \centering
  \includegraphics[width=.85\textwidth]{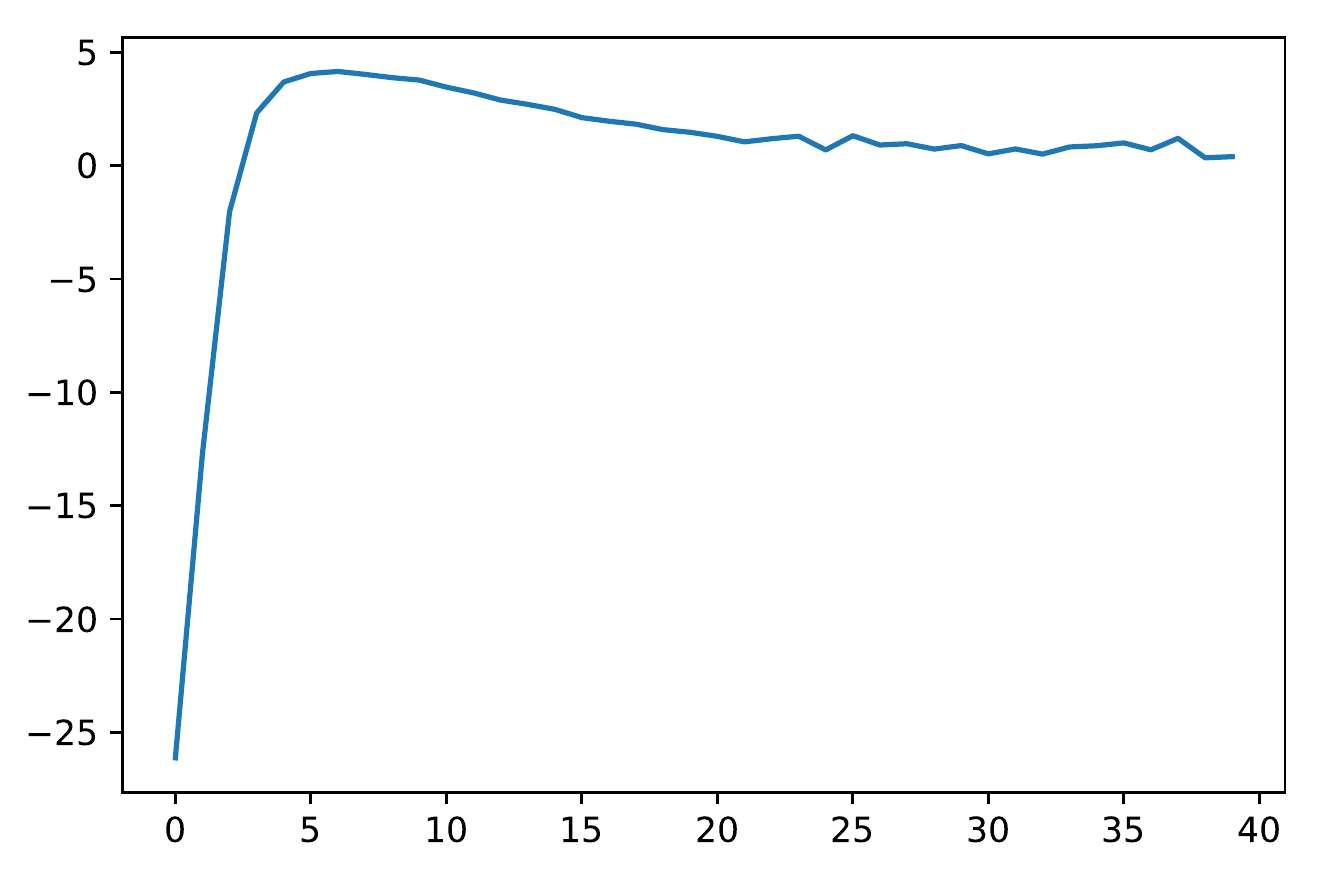}
\end{minipage}%
\begin{minipage}{.5\textwidth}
  \centering
  \includegraphics[width=.85\textwidth]{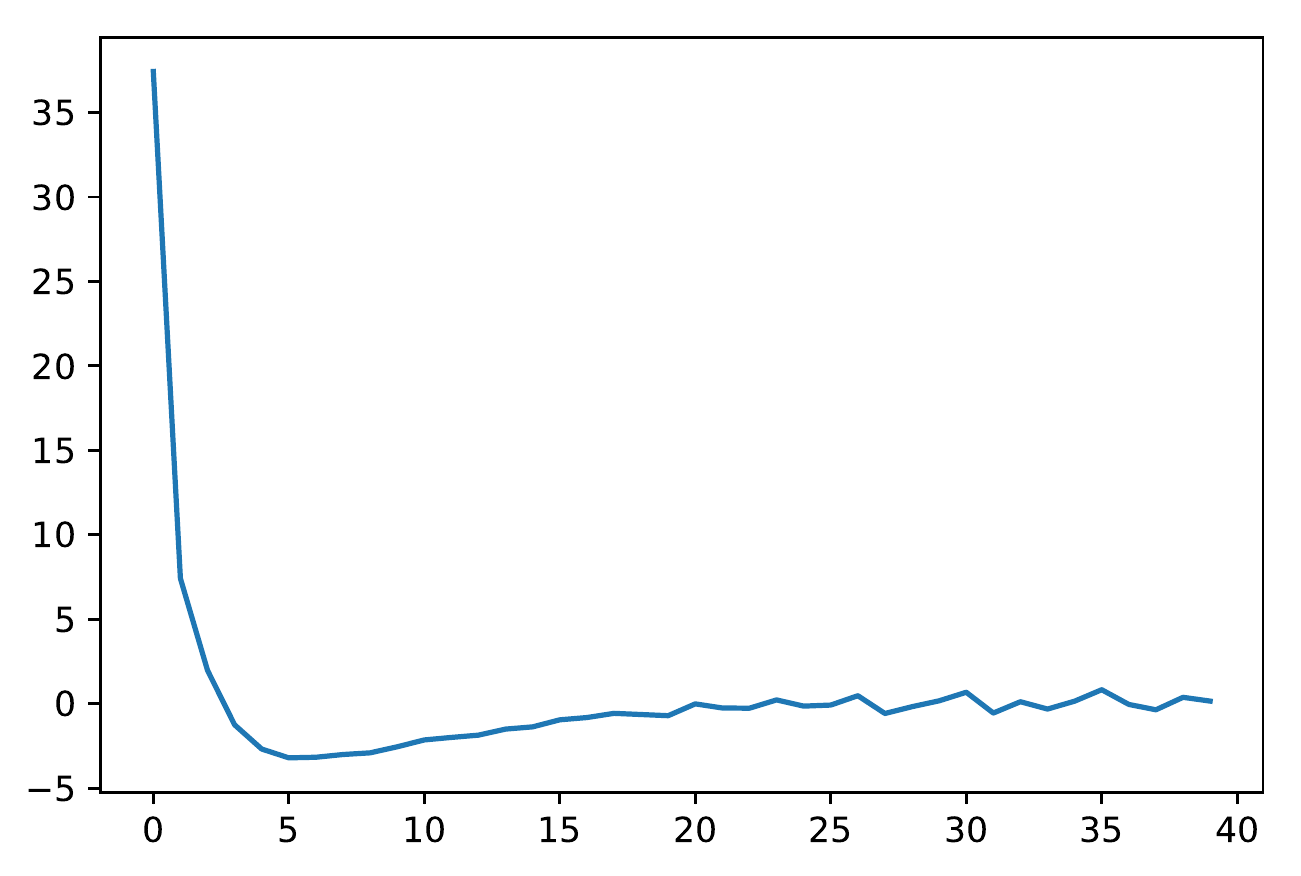}
\end{minipage}
\caption{Empirical evaluation of (c) of Section \ref{subsec:backtest} for the sets 
$B_1$ (left) and $B_2$ (right) under $\p \otimes \q_E$.}
\label{VActest}
\end{figure}

To test (c) from Section \ref{subsec:backtest} we considered the sets 
\[
B_1 = \crl{x \in \R^3 : x_1 > q_{70\%} \mbox{ and } x_2 < r_{30\%}}
\quad \mbox{and} \quad
B_2 = \crl{x \in \R^3 : x_1 < q_{30\%} \mbox{ and } x_2 > r_{70\%}}
\]
where $q_{\beta}$ and $r_{\beta}$ denote the $\beta$-quantiles of $q_{\tau}$ and $r_{\tau}$ 
under $\p \otimes \q_E$; see Figure \ref{VActest}.

\begin{figure}[ht!]
\centering
\begin{minipage}{.5\textwidth}
  \centering
  \includegraphics[width=.85\textwidth]{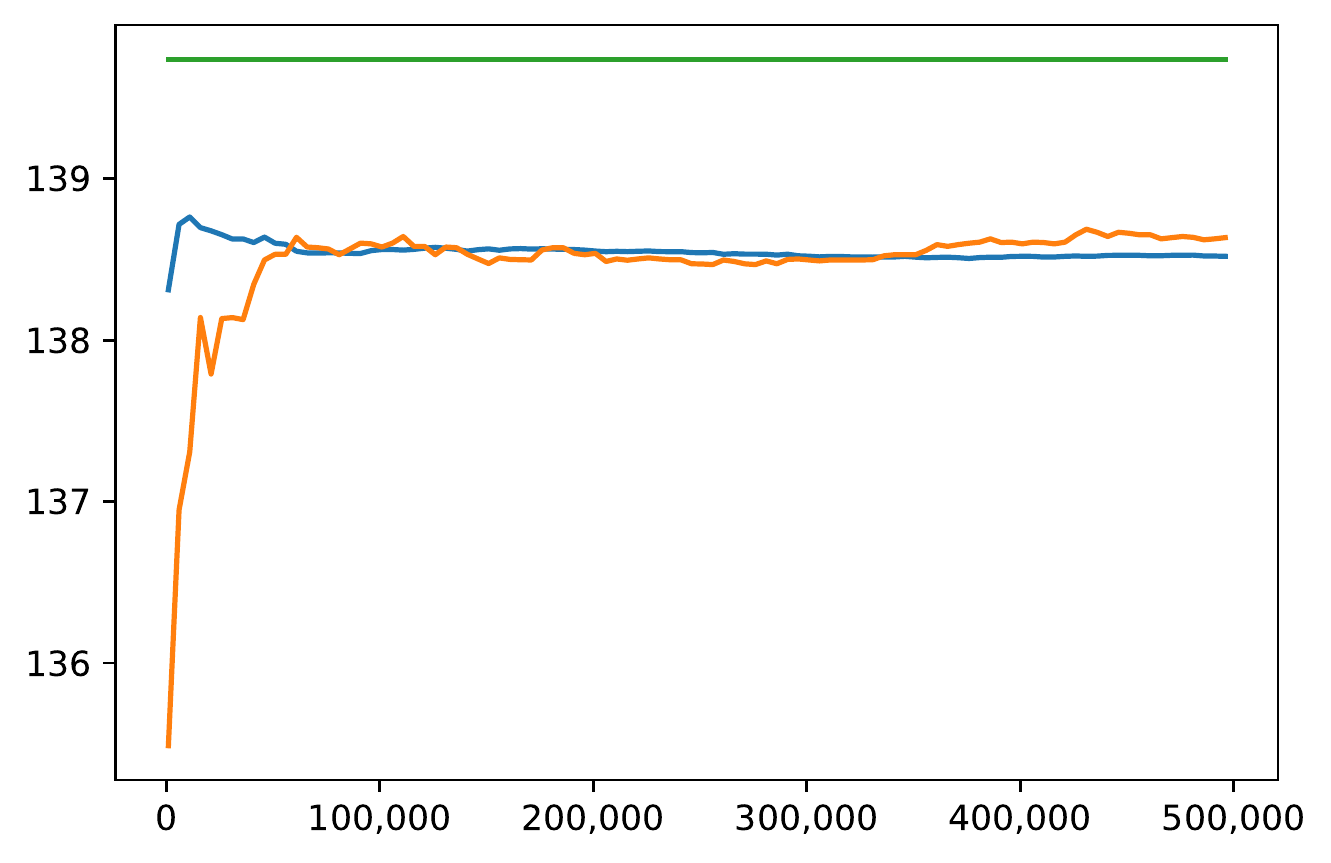}
 \end{minipage}%
\begin{minipage}{.5\textwidth}
  \centering
  \includegraphics[width=.85\textwidth]{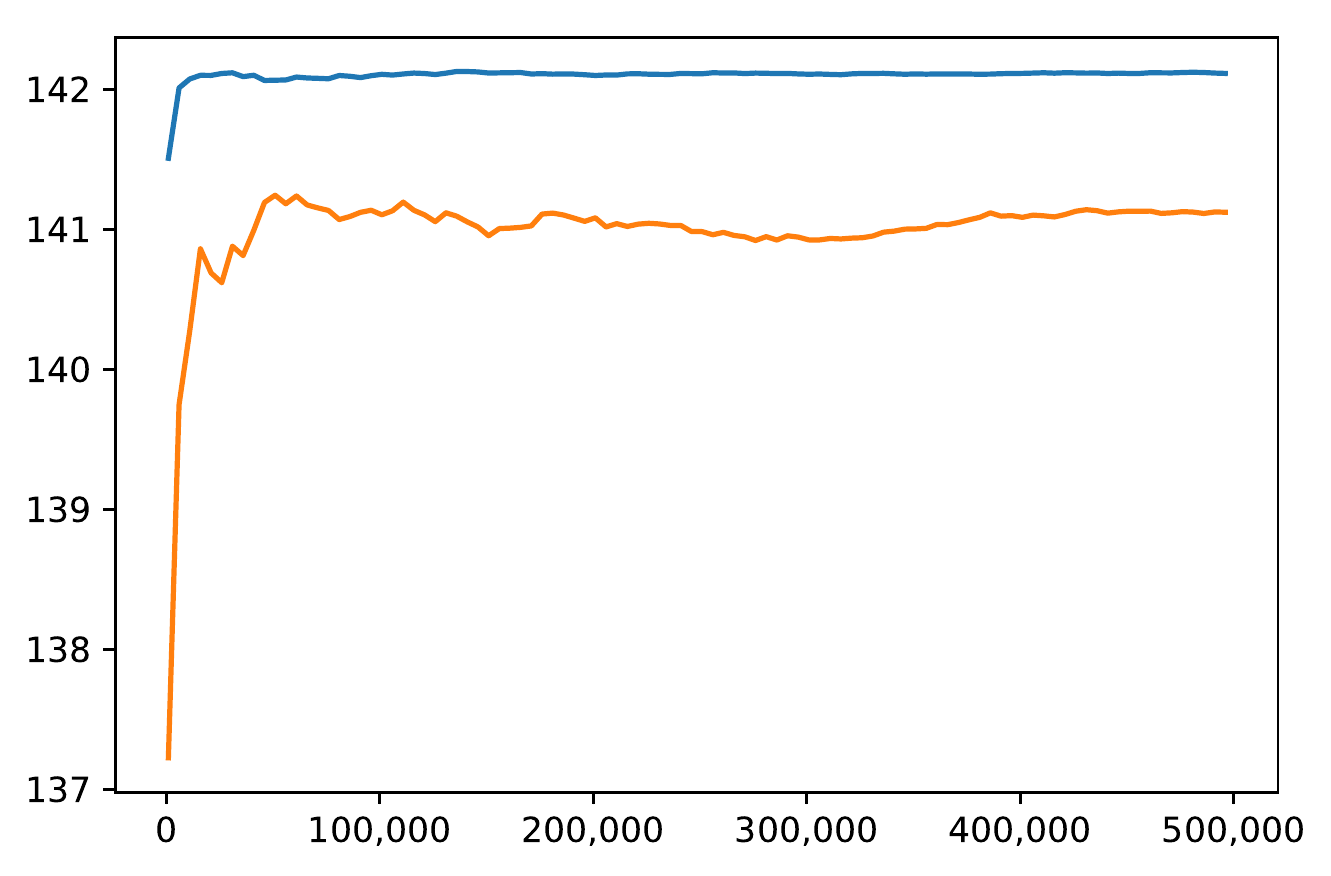}
\end{minipage}
\caption{Convergence of the empirical 99.5\%-VaR (left) and 99\%-ES (right) without IS (orange) and with IS (blue) 
compared to the reference value from \cite{bauer} (green).} \label{fig:VArange}
\end{figure}

We also trained and tested under $\p^{\nu} \otimes \q_E$ for an IS distribution $\nu$ on $\R^d$.
Our results are reported in Figure \ref{fig:VArange}.
Our estimate of $\mbox{VaR}_{99.5\%}(L)$ was 138.64 without and 138.52 with IS.
In comparison, the VaR estimate obtained by \cite{bauer} using 37 monomials and 40 million 
simulations is 139.74. Our estimates of $\mbox{ES}_{99\%}(L)$ came out as 
141.12 without and 142.12 with IS. There exist no reference values for this case.

\section{Conclusion}\label{sec:conclu}

In this paper we have developed a deep learning method for 
assessing the risk of an asset-liability portfolio over a given time horizon.
It first computes a neural network approximation of the portfolio value at 
the risk horizon. Then the approximation is used to estimate a risk measure, 
such as value-at-risk or expected shortfall from Monte Carlo simulations. 
We have investigated how to choose the architecture of the 
network, how to learn the network parameters under a suitable importance sampling distribution
and how to test the adequacy of the network approximation. We have illustrated the approach
by computing value-at-risk and expected shortfall in three typical risk assessment problems
from banking and insurance. In all cases the approach has worked efficiently and 
produced accurate results. 

\bibliographystyle{apa}

\end{document}